\documentclass[iop]{emulateapj}
\usepackage{natbib,graphicx,amsmath,xspace}
\usepackage{color}

\newcommand{\grs}{\mbox{\object{GRS 1915+105}}\xspace}

\newcommand{\nustar}{\emph{NuSTAR}\xspace}
\newcommand{\chandra}{\emph{Chandra}\xspace}
\newcommand{\rxte}{\emph{RXTE}\xspace}
\newcommand{\kms}{\mbox{\xspace km s$^{-1}$\xspace}}

\begin{document}

\title{Disk-Wind Connection during the Heartbeats of \grs}

\author{Abderahmen ~Zoghbi$^{1}$, J.~M.~Miller$^{1}$, A.~L.~King$^{2}$, M. C. Miller$^{10}$, D.~Proga$^{3}$, T.~Kallman$^{4}$, A.~C.~Fabian$^5$, F.~A.~Harrison$^6$, J. Kaastra$^{7,8}$, J. Raymond$^{9}$, C.~S.~Reynolds$^{10}$, S.~E.~Boggs$^{11}$, F.~E.~Christensen$^{12}$, W.~Craig$^{11}$, C.~J.~Hailey$^{13}$,  D.~Stern$^{14}$, W.~W.~Zhang$^4$}
\affil{$^1$Department of Astronomy, University of Michigan, Ann Arbor, MI 48109, USA}
\affil{$^2$KIPAC, Stanford University, 452 Lomita Mall, Stanford, CA 94305, USA}
\affil{$^3$Department of Physics, University of Nevada, Las Vegas, Las Vegas, NV 89154, USA}
\affil{$^4$NASA Goddard Space Flight Center, Code 662, Greedbelt, MD
20771, USA}
\affil{$^5$Institute of Astronomy, University of Cambridge, Madingley Road,
Cambridge CB3 OHA, UK}
\affil{$^6$Space Radiation Laboratory, California Institute of Technology, Pasadena, CA 91125, USA}
\affil{$^7$SRON Netherlands Institute for Space Research, Sorbonnelaan 2,
3584 CA Utrecht, NL}
\affil{$^8$Department of Physics and Astronomy, Universiteit Utrecht, PO Box
80000, 3508 TA Utrecht, NL}
\affil{$^9$Harvard-Smithsonian Center for Astrophysics, 60 Garden Street,
Cambridge, MA 02138, USA}
\affil{$^{10}$Department of Astronomy, University of Maryland, College Park,
MD 20742-2421, USA}
\affil{$^{11}$Space Science Laboratory, University of California, Berkeley, California 94720, USA}
\affil{$^{12}$DTU Space. National Space Institute, Technical University of Denmark, Elektrovej 327, 2800 Lyngby, Denmark}
\affil{$^{13}$Columbia Astrophysics Laboratory, Columbia University, New York, New York 10027, USA}
\affil{$^{14}$Jet Propulsion Laboratory, California Institute of Technology, Pasadena, CA 91109, USA}
\email{abzoghbi@umich.edu}

\begin{abstract}
Disk and wind signatures are seen in the soft state of Galactic black holes, while the jet is seen in the hard state.
Here we study the disk-wind connection in the $\rho$ class of variability in \grs using a joint \nustar-\chandra observation. The source shows 50 sec limit cycle oscillations. By including new information provided by the reflection spectrum, and using phase-resolved spectroscopy, we find that
the change in the inner disk inferred from the blackbody emission is not matched by reflection measurements. The latter is almost constant, independent of the continuum model. The two radii are comparable only if the disk temperature color correction factor changes, an effect that could be due to the changing opacity of the disk caused by changes in metal abundances. 
The disk inclination is similar to that inferred from the jet axis, and oscillates by $\sim10^{\circ}$.
The simultaneous \chandra data show the presence of two wind components with velocities between 500-5000 \kms, and possibly two more with velocities reaching $20,000$ \kms ($\sim 0.06\,c$). The column densities are $\sim5\times10^{22}\, {\rm cm}^{-2}$. An upper limit to the wind response time of 2 sec is measured, implying a launch radius of $<6\times10^{10}\, {\rm cm}$. The changes in wind velocity and absorbed flux require the geometry of the wind to change during the oscillations, constraining the wind to be launched from a distance of $290-1300\, r_g$ from the black hole. Both datasets support fundamental model predictions in which a bulge originates in the inner disk and moves outward as the instability progresses.
\end{abstract}

\section{Introduction}
\grs is a microquasar with some extreme properties. Its unique X-ray variability classes \citep{2000A&A...355..271B}, the superluminal radio jets \citep{1994Natur.371...46M,2004ARA&A..42..317F} and the observation of absorption lines from possible winds \citep{2002ApJ...567.1102L} make it a unique laboratory for testing accretion physics as well as investigating questions on how the disk, jet, and wind are linked. The convenience of the short time-scales of these interactions is extremely valuable in understanding the analogous processes taking place at galactic scales and beyond.

\grs was discovered by \emph{GRANAT} \citep{1994ApJS...92..469C} as a 300 mCrab transient.
It was the first object to show superluminal jet motion in the Galaxy \citep{1994Natur.371...46M}.
Initial distance estimates put it at 12.5 kpc \citep{1994Natur.371...46M}, but more recent parallax measurements suggest the distance to be $8\pm2$ kpc \citep{2014ApJ...796....2R}.
The binary orbital period is $\sim33.5$ days and the black hole mass is likely to be $10.1\pm0.6$ solar masses \citep{2013ApJ...768..185S}.

With the launch of \emph{RXTE}, the glory of \grs was uncovered. Complex and structured light curves have been attributed to disk instabilities from the start \citep{1996ApJ...473L.107G,1997ApJ...488L.109B}. 
\cite{2000A&A...355..271B} presented a classification of the variability classes of \grs. With the additional hardness ratio information, the complex variability was empirically interpreted as transitions between three \emph{spectral} states, named A, B and C. A and B both show soft spectra and correspond to stable high flux periods, with B having higher temperature and stronger red noise variability. State C shows hard colors with the spectra dominated by a flat powerlaw. The complexity of the variability can be reduced to transition between these states on different time-scales. The link between these states and the canonical black hole binary states \citep{2006ARA&A..44...49R} is not clear, but state C shares many properties with the hard state, while A and B have the properties of the steep powerlaw state. The $\rho$ {\emph variability} class is characterized by coherent oscillations with periods of 50--100 seconds between the low C spectral state and the high A/B states.

The general picture acquired mostly though Proportional Counter Array (PCA) data from \emph{RXTE} is that a thermal-viscous instability \citep{1974ApJ...187L...1L} of the standard \cite{1973A&A....24..337S} disk is responsible for the transitions between the different spectral states, producing the structured light curves and the limit cycle behavior. The broad band spectra in the \emph{RXTE} era were modeled as a thermal multicolor blackbody with a hard powerlaw (with a cutoff) and an emission line at 6-7 keV \citep[e.g.][]{2004ARA&A..42..317F}. The existence of a high energy cutoff depends strongly on the assumed model.

During the structured oscillations, the inner radius of the disk inferred from modeling the blackbody emission appears to change. This suggested that the central region of the disk evaporates or becomes radiatively inefficient during the instability before refilling on a viscous time scale. This picture appeared to be supported by the spectral modeling of the different states \citep{1997ApJ...488L.109B,2000A&A...355..271B,2004ARA&A..42..317F,2012A&A...537A..18M}. Uncertainties in disk atmosphere models (e.g. the density or the fraction of power dissipated in a corona) can however be important and may affect these inferred radius estimates significantly \citep[e.g.][]{2000MNRAS.313..193M,2009ApJ...697..900M}.

In terms of the accretion physics, the Shakura-Sunyaev disk is unstable at high mass accretion rates if the viscous stress is proportional to the total, gas plus radiation pressure \citep{1973A&A....29..179P,1974ApJ...187L...1L}. This is caused by the increase in heating as radiation pressure becomes significant compared to gas pressure at high accretion rates. The strong dependence of radiation pressure on temperature leads to runaway heating not compensated by a decrease in opacity, and the disk becomes unstable. Stability can be provided by additional cooling provided, for example, by advection \citep{1988ApJ...332..646A} or if a significant fraction of the accretion energy is channeled away from the disk to power a corona \citep{2006MNRAS.372..728M,2016arXiv160106785S}. Observationally, disks in the thermal state are stable up to 0.7 Eddington, while the expectation is that the instability kicks in at much lower accretion rates. An $\alpha$-prescription different from the simple $P_g+P_{rad}$ (e.g. geometric mean of the two) might explain the limit cycle oscillation of \grs \citep{2006MNRAS.372..728M,2007A&ARv..15....1D}.

The question of disk stability in numerically simulated accretion disks is an active topic of research. There is reason to expect the heating due to the magneto-rotational instability to be proportional to the total pressure, and by using shearing box simulations that incorporate radiation transport and cooling, the question of stability can be studied \citep[e.g.][for a recent review]{2014SSRv..183...21B}. Some of the early simulations \citep{2004ApJ...605L..45T,2009ApJ...704..781H} of disks dominated by radiation pressure were stable for many thermal times. More recent simulations on the other hand \citep{2013ApJ...767..148J} suggest that such disks eventually suffer runaway heating or cooling depending on disk properties. Additionally, the radial flux of mass and energy could be important during the instability, and are not captured in the limited shearing box simulations. More observational constraints are therefore very important.

For outflows, absorption lines in the iron band in \grs were first seen in \emph{ASCA} data \citep{2000ApJ...539..413K} and later confirmed with \chandra/HETGS \citep{2002ApJ...567.1102L,2011ApJ...737...69N}. The behavior of the jet radio emission in relation to the X-ray states is complex, but several clear trends have been observed \citep{2004MNRAS.355.1105F}. The low flux C state (or canonical hard states in other objects) is associated with a steady jet emission. Prolonged A/B states are associated with jet suppression and optically thin jets are ejected during the transition between C and A states. Observations show that winds and jets are anti-correlated \citep{2006Natur.441..953M,2009Natur.458..481N,2012ApJ...746L..20K,2012MNRAS.422L..11P}, suggesting that the changes in the accretion physics are tightly linked to the outflow mechanism \citep[e.g.][]{2015ApJ...809..118B}.

All previous work on the detailed behavior of \grs has relied on PCA data mostly because of their large collecting area. However, given the limited energy resolution, studying the more powerful reflection spectrum had to wait until \nustar \citep{2013ApJ...770..103H} was launched, which is the ideal instrument since it has the timing capabilities, sensitivity and energy resolution to study the relativistic reflection during the extreme variability from the inner regions for the first time. Here, we present phase-resolved spectroscopy of oscillations in the $\rho$-state, or the heartbeats state of \grs, characterized by a strong coherent oscillations with a period of $\sim50$ sec.

\begin{figure}
\centering
 \includegraphics[width=210pt,clip ]{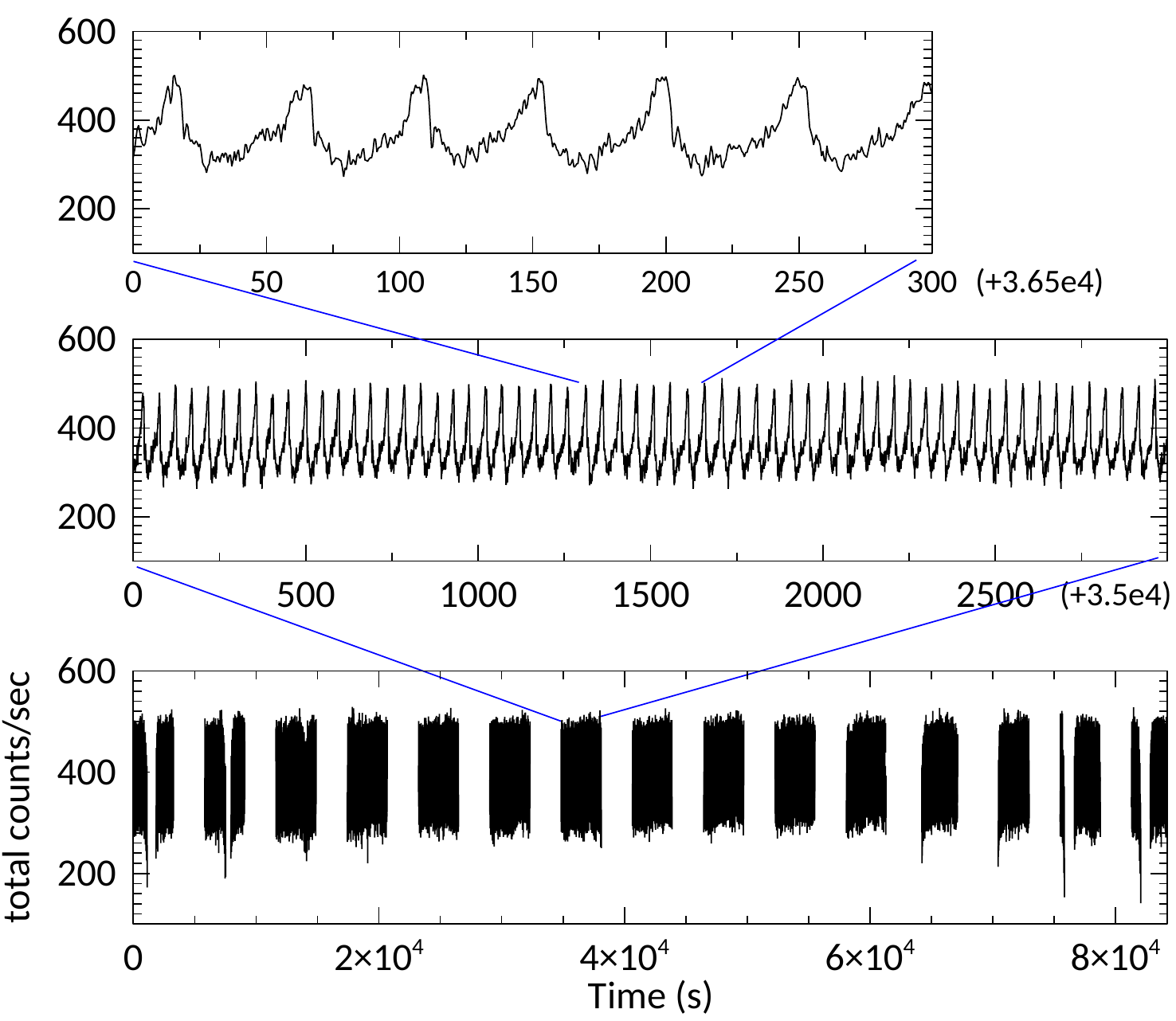}
\caption{3--79 keV light curve of GRS 1915+105 from \nustar. The light curve was extracted with a time bin of 0.03125 seconds and then smoothed with a boxcar function with 1 second width. The two upper panels show a zoom-in on the indicated parts of the light curve.}
\label{fig:lc_all}
\end{figure}

\section{Observations \& Data Reduction}
\nustar observed \grs starting February 23, 2015 for 40 ks (ID: 90001001002). The data were reduced following the standard procedure using the relevant software in \textsc{heasoft} 6.17 and using the calibration files from CALDB v. 20150702. A light curve from the whole observation is shown in Figure \ref{fig:lc_all}. These were extracted from a 2.5 arcmin radius circular region centered at the source. The time sampling of the light curves is 0.03125 sec. The plotted curves have been smoothed with a running mean box function of length 32 (i.e. 1 sec) for display clarity. The light curve shows the source in the $\rho$ state as defined in \cite{2000A&A...355..271B} with stable oscillations of $\sim 50$ sec throughout the observation.

The source spectra in the following discussions have been extracted using \texttt{nuproducts} from circular regions of radius 2.5 arcmin centered on the source. Background spectra are extracted from a source-free region of a similar size. The spectral fitting was done in \textsc{xspec version 12.9.0} using $\chi^2$ statistics after the spectra were grouped to have a minimum signal to noise ratio per bin of 6 and a minimum energy separation equal to 0.3 the instrument resolution. Spectra from the two \nustar modules were combined before doing the spectral modeling. Modeling spectra from the two modules separately gives similar results.

\section{Average Spectra}\label{sec:avg_spec}
We start by analysing the average spectrum of \grs from the \nustar observation. In the following spectral modeling, Galactic absorption is modeled with \texttt{tbabs} \citep{2000ApJ...542..914W}. The value of the column density was estimated using the simultaneous \chandra data (discussed in detail in section \ref{sec:chandra_data}), where edge models are fitted individually to the photoelectric edges from S, Si and Fe using the High Energy Grating Spectrometer (HEGS), and to Mg using the Medium Energy Grating Spectrometer (MEGS). The equivalent Hydrogen column densites from these elements range between $3-7\times10^{22}\,{\rm cm}^{-2}$, with significant differences, suggesting non-solar abundances \citep[][though our estimates are slightly different]{2002ApJ...567.1102L}. The column density in the \nustar data was therefore estimated directly during the spectral modeling of the average spectra. The best fit value is $N_{\rm H} = 4.9\pm0.1\times10^{22}\, {\rm cm}^{-2}$.

\begin{figure}
\centering
 \includegraphics[width=200pt,clip ]{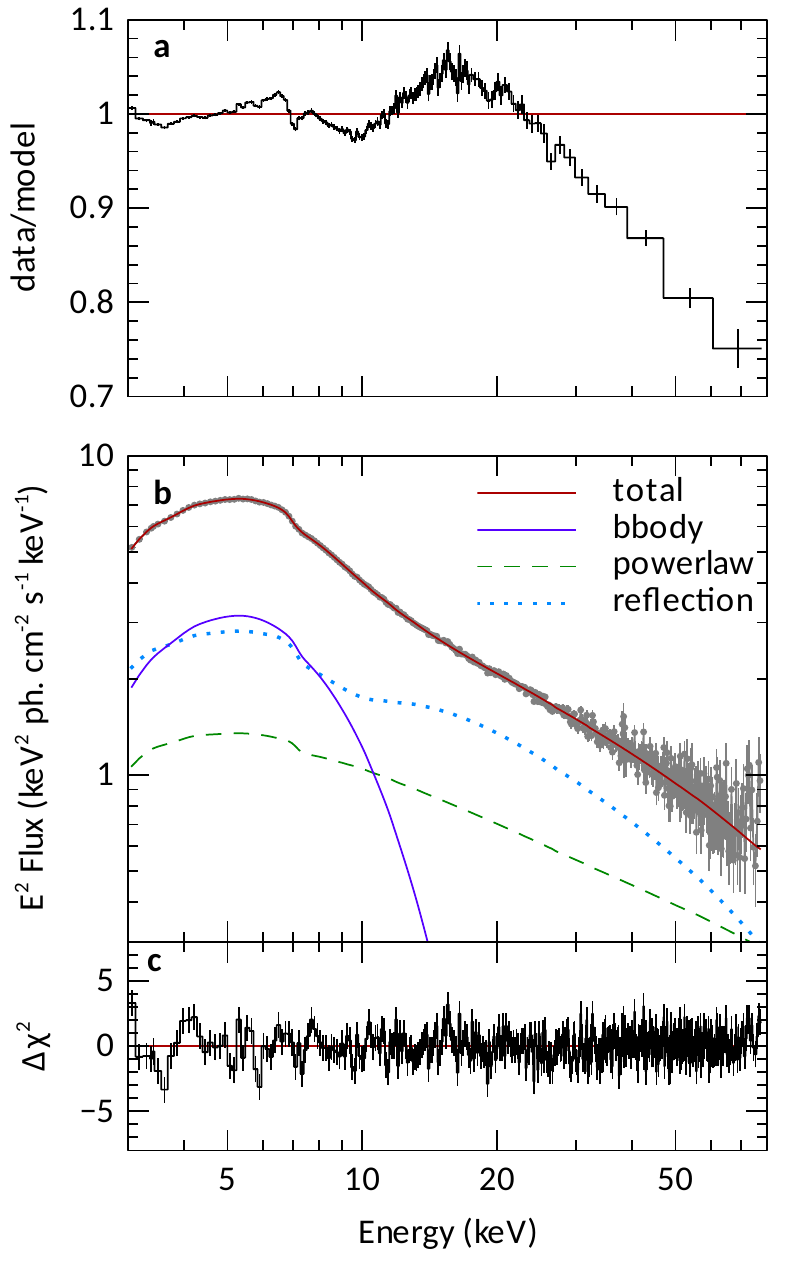}
\caption{{\bf a:} Ratio of the average spectrum to a model consisting of a multicolor disk blackbody \texttt{ezdiskbb} and a \texttt{powerlaw}. {\bf b:} The average spectrum and its best fitting model. The model has the \textsc{xspec} form: \texttt{tbabs * (ezdiskbb + relxill + cutoffpl + gaussian)} (see text for details). The effective area of the detector has been removed by unfolding the spectrum against a constant.}
\label{fig:nu_fit_avg}
\end{figure}

The average \nustar spectrum has a significant contribution from a strong soft component, most likely a thermal disk component in addition to the hard power-law-like component. Panel (a) in Figure \ref{fig:nu_fit_avg} shows the ratio of the spectrum to a model consisting of a powerlaw and a blackhody disk \citep[\texttt{ezdiskbb};][]{2005ApJ...618..832Z}. Strong residuals are visible, particularly at the iron energies and a peak at $\sim 20$ keV, indicating the existence of a strong reflection component. An absorption line is visible at 7 keV, likely due to Fe {\sc xxvi}. There are weak residuals around 5 keV that could be absorption from elements such as Ca, Cr and Mn, but they are not very significant and we do not discuss them further.

A reflection component is added to account for these broad emission features. We use the \texttt{relxill} model \citep{2010MNRAS.409.1534D,2014ApJ...782...76G}. The model calculates the reflection spectrum from a disk illuminated by a bright powerlaw X-ray source. The spectrum is then convolved with a general relativistic kernel to account for the Doppler broadening and relativistic redshift resulting from the black hole gravitational potential. The absorption line at 7 keV is modeled with a Gaussian line. The detailed modeling of this feature is deferred to the analysis of the \chandra data in section \ref{sec:chandra_data}. The model therefore has the \textsc{xspec} form: \texttt{tbabs*(gaussian+ezdiskbb+relxill+cutoffpo)}. The \texttt{relxill} model was used to fit the reflection spectrum only and the illuminating powerlaw is included as a separate component. The normalization of the Gaussian line is negative in this case to model the absorption line. The Gaussian line was consistent with being unresolved, so its width was fixed at zero. This model gives a reasonably good description of the data with a reduced $\chi^2$ goodness of fit of 1.2 for 422 degrees of freedom. The best fit parameters are summarized in Table \ref{tab:avg_spec}. 

The reflection spectrum in the average spectrum is consistent with the accretion disk extending down to the inner most stable circular orbit (ISCO) of a highly spinning black hole ($a = 0.993\pm0.05$), in agreement with the first \nustar observation \citep{2013ApJ...775L..45M}. The negative $q_2$ suggests that the disk is not flat during the oscillation, a point we discuss further in section \ref{sec:nu_phase} The inner radius ($R_{\rm bb}$) can also be estimated from the blackbody spectrum \citep{1997ApJ...482L.155Z}, and it is related to the multicolor blackbody normalization by \citep{2005ApJ...618..832Z}:
\begin{equation}\label{eq:bb}
N_{\rm bb} = \frac{1}{f_{c}^4} \Big(\frac{R_{\rm bb}}{D}\Big)^2 {\rm cos}\, i
\end{equation}

where $f_c$ is the color correction factor \citep{1995ApJ...445..780S}, and $D$ is the black hole distance in units of 10 kpc. Assuming a distance of $D=8\pm2$ kpc \citep{2014ApJ...796....2R} and $f_c=1.7$ and $i=67^{\circ}$ as inferred here \citep[see also][]{1999MNRAS.304..865F}, we find $R_{\rm bb}=17\pm2$ km. The inner radius measured from fitting the reflection spectrum is $R_{\rm ref}=20\pm1$ km, assuming a black hole mass of $M=10.1\pm0.6M_{\odot}$ \citep{2013ApJ...768..185S}. The two values are statistically equivalent. Also, note that the flux of the blackbody does not include photons intercepted by the Comptonizing medium. The true blackbody flux is therefore higher, implying implying a slightly larger value for $R_{\rm bb}$. If we model the Comptonization with \texttt{simpl} \citep{2009PASP..121.1279S}, which conserves the number of photons, we obtain $R_{\rm bb}=22\pm3$ km, which is even more consistent with $R_{\rm ref}$. Further discussion of the radius measurements is presented in section \ref{sec:nu_phase}. Although we used a broken powerlaw emissivity model in Table \ref{tab:avg_spec} (with $q_2$ being the index above the the break radius at $r_{\rm br}$), exploring the uncertainties of $q_2$ and $r_{\rm br}$ with Monte Carlo Markov Chains suggests that they are poorly constrained. Although a single powerlaw emissivity gives a comparable quality fit, we use a broken powerlaw because it is more theoretically motivated \citep{2012MNRAS.424.1284W}. This result is a consequence of the fact that the line is very broad with little emission at the core of the line that would constrain emission from outer radii.

\begin{table}
\centering
\begin{tabular}{|lll|}
\hline
\texttt{ezdiskbb}&&\\
$N_{\rm bb}$: $17\pm1$ & $T_{\rm bb}$: $1.612\pm0.003$ & \\
\hline
\texttt{cutoffpo}&&\\
$N_{\rm po}$: $5.2\pm1.2$&$\Gamma$: $2.65\pm0.04$&$E_{\rm c}$: $>976$ keV\\
\hline
\texttt{relxill}&&\\
$N_{\rm ref}$: $0.35\pm0.01$ & $\theta$: $67.2\pm0.2^{\circ}$&$r_{\rm in}$: $1.33\pm0.01\ r_g$\\
$q_1$: $9.7\pm0.3$ & $q_2$: $<0$&($a = 0.993\pm0.05$)\\
$log(\xi)$: $3.98\pm0.06$& $A_{\rm Fe}$: $0.54\pm0.03$&$r_{\rm br}$: $50\pm23\ r_g$\\
\hline
\texttt{gaussian}&&\\
$E_{\rm abs}$: $7.06\pm0.01$ (keV)& \multicolumn{2}{l|}{$N_{\rm abs}$: $(-1.73\pm0.01)\times10^{-3}$}\\
\hline
\end{tabular}

\caption{Best fit parameters to the average spectrum and their $1\sigma$ statistical uncertainties. The model has the \textsc{xspec} form: \texttt{tbabs * (ezdiskbb + relxill + cutoffpl + gaussian)} (see text for details). $N$ stands for normalization.}
\label{tab:avg_spec}
\end{table}

\section{Phase-Resolved Spectra}
In order to probe how the components change during the coherent oscillations seen in Figure \ref{fig:lc_all}, we extract spectra as a function of the phase of the oscillations. We use a modified version of the method of \cite{2011ApJ...737...69N} to obtain the waveform of the periodic oscillation and Good Time Intervals (GTI) for phase bins.

\subsection{\nustar}\label{sec:nu_phase}
\subsubsection{Spectral Extraction}
We extract a light curve with time bins of $2^{-5}$ sec and apply barycenter corrections. We use the whole \nustar energy band (3-79 keV) and we sum light curves from modules A and B. A starting approximate template for the waveform is moved across the light curve, and a measure of the match between the template and the light curve at every point is calculated. This is similar to cross-correlation, but we use the sum of the squared difference as a match estimator. We found this to change smoothly across the light curve allowing easier identification of minima. The minima in the resulting match estimator give the positions of light curve segments that best match the template. Using this method takes into account the quasi-periodic nature of the oscillations and the random phase changes. A new template is constructed by averaging the resulting segment matches, and the process is repeated a few times until the resulting waveform ceases to change. From the positions of the matching segments, we take the peak-to-peak time difference to correspond to the phase range $0-2\pi$. Hence, for every time bin in the light curve, we associate a phase that corresponds to the position of the bin relative to the neighboring oscillation peaks. No phase information is associated with time bins where the oscillation and the waveform shape are not well defined. GTIs are then constructed by selecting light curve segments that fall in the phase bin of interest. Spectra are subsequently extracted by passing these GTIs to \texttt{nuproducts}.

\begin{figure}
\centering
 \includegraphics[width=240pt,clip ]{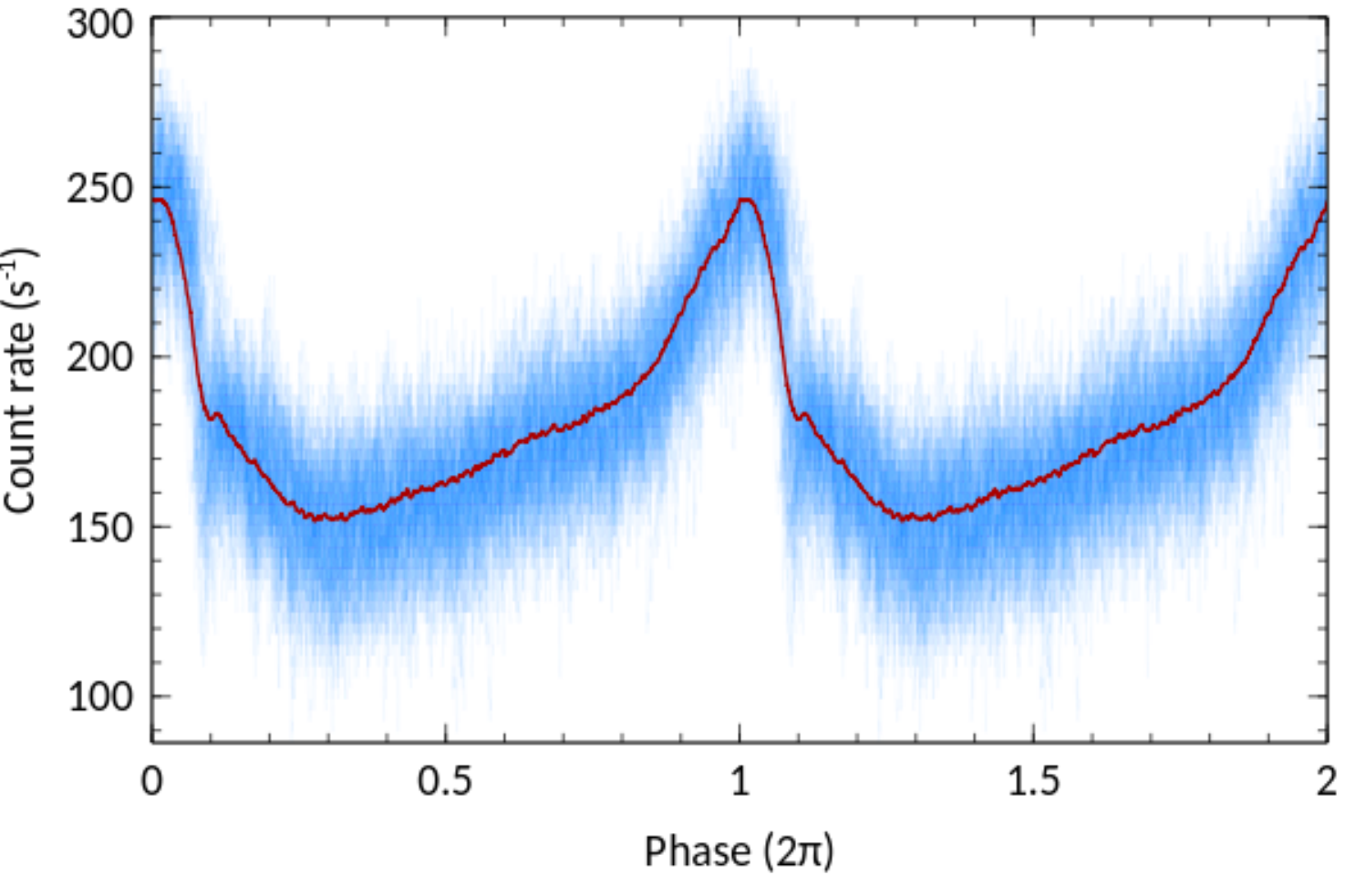}
\caption{The average oscillation waveform from the light curve. The thin (blue) lines represent a sample of light curve segments and the thick (red) line is the average waveform.}
\label{fig:nu_waveform}
\end{figure}

The resulting waveform of the oscillation is shown in Figure \ref{fig:nu_waveform}. The plot covers two repeated periods to clearly show the shape of the oscillation. The blue color shows the waveform from all the segments, while the red line shows the average. The oscillation is characterized by a slow rise from the minimum (0.3--0.8) followed by an increase in the rate of change (0.8--1.0) and then a fast drop. In the following discussion, we will refer to the phases in units of $2\pi$, so the range of oscillation phase is between 0 and 1. 

\begin{figure*}
\centering
 \includegraphics[width=480pt,clip ]{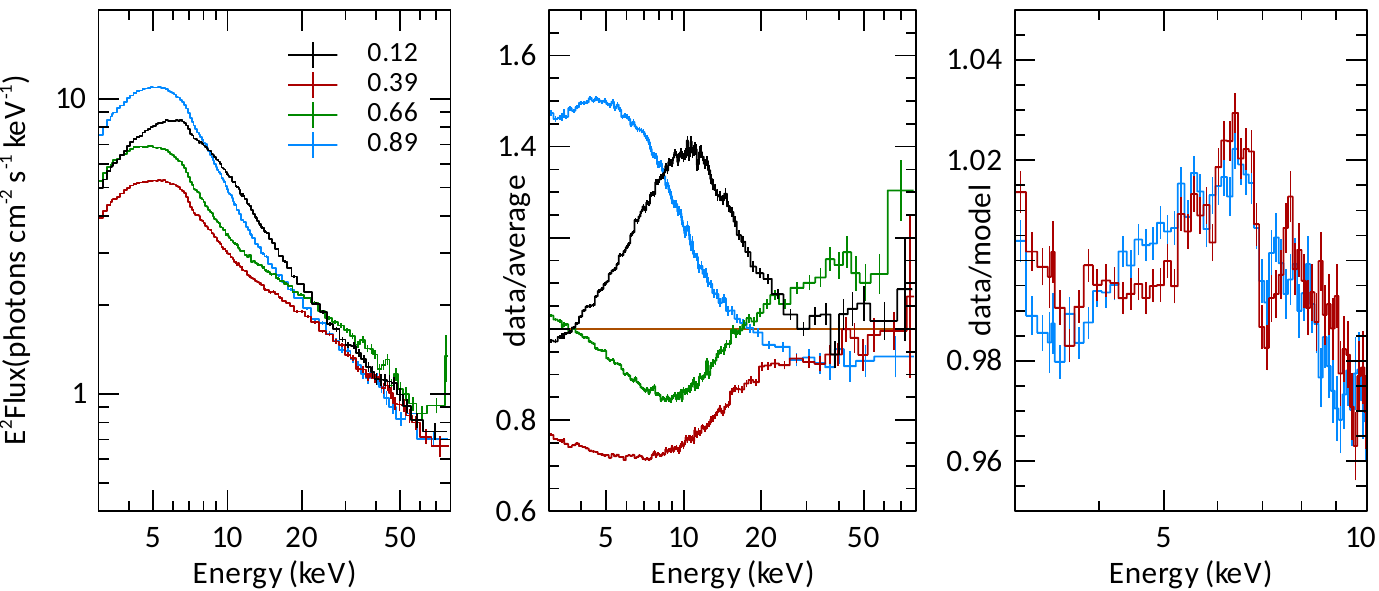}
\caption{\emph{Left:} Phase-resolved spectra at the four phases indicated. The phase bin widths are 0.12, 0.15, 0.13 and 0.11 for the four bins shown respectively. The spectra have been divided by the effective area of the detector by unfolding them through a constant model. \emph{Middle:} The same four spectra from the left panel, but now plotted as a ratio to the average spectrum. \emph{Right:} Phase spectra at two phases showing a zoom-in at the broad iron line at 6.4 keV.}
\label{fig:nu_fit0_show_spec}
\end{figure*}

For an initial look at the spectral changes with phase, Figure \ref{fig:nu_fit0_show_spec} presents the spectra from four phase bins, where the bins are selected so that individual spectra have roughly equal numbers of total counts ($\sim4\times10^{6}$ in this case). Figure \ref{fig:nu_fit0_show_spec} shows three model-independent representations of the spectra. The left panel shows the spectra after correcting for the effective area of the detector (by unfolding the spectra to a constant model). The middle panel shows the ratio of individual spectra to the average total spectrum shown in Figure \ref{fig:nu_fit_avg}-b, and the right panel shows a zoom-in view of the iron line energies plotted as a ratio to a (\texttt{powerlaw+ezdiskbb}) model.

The left and middle panels show that the main spectral changes happen at soft energies; these can be seen as both a change in amplitude of the soft component (e.g. $\phi=0.39$ vs $\phi=0.89$) as well as a shift in energy ($\phi=0.12$ vs $\phi=0.89$). Changes also happen at energies greater than 20 keV and can be seen as simple flux changes. These changes are better explored by fitting models to the spectra and tracking their variability. The right panel of Figure \ref{fig:nu_fit0_show_spec} shows the changes in the relativistic iron line for two representative phases. As can be seen, the line shape appears to change with phase: the broad iron line at $\phi=0.39$ appears somewhat narrower compared to the $\phi=0.89$ phase.

\subsubsection{Phenomenological Modeling}
To explore the details of the spectral variability, we fit the phase spectra with several models and explore changes in the model parameters. We start with a simple model consisting of a multicolor disk component (\texttt{ezdiskbb}) and a cutoff powerlaw to model the hard energy emission. This is a phenomenological model and serves as a direct comparison with the similar modeling in \cite{2011ApJ...737...69N}. All the following models are absorbed by a constant column of $N_h=4.9\times10^{22} \rm{cm}^{-2}$. Here, we extract the spectra from 10 phase bins chosen to have roughly equal total numbers of net counts ($\sim 1.5\times10^6$ counts per phase bin).

\begin{figure}
\centering
 \includegraphics[width=200pt,clip ]{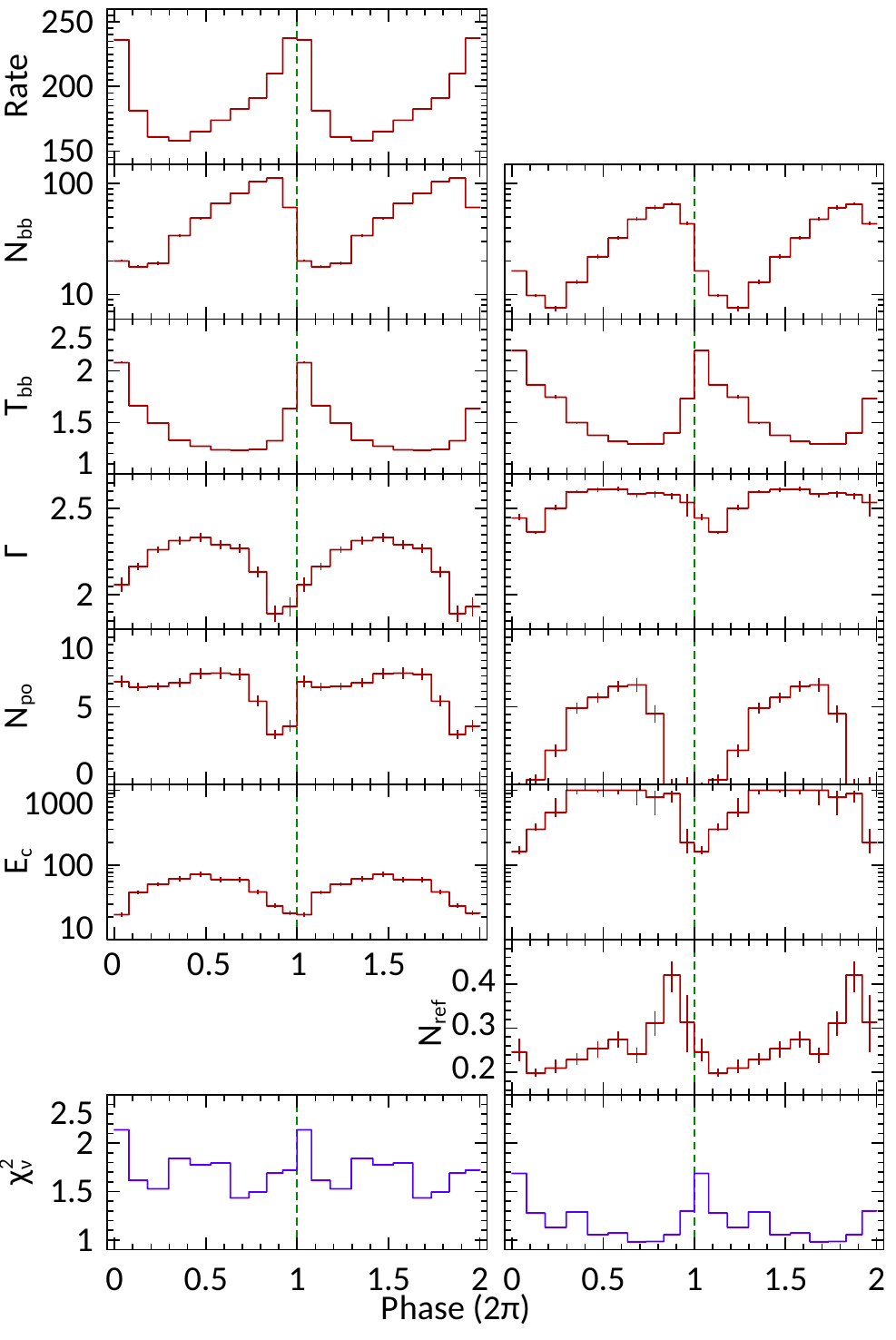}
\caption{Parameter fits from modeling the phase-resolved spectra. The horizontal axis in all panels is the phase of the oscillation in units of $2\pi$. The plots from phases 0 to 1 are repeated to phases 1 to 2 to show the oscillations more clearly. The top panel on the left shows the total count rate oscillation and the bottom two panels show the $\chi^2$ goodness of fit statistics from the spectral modeling. \emph{Left Column:} Parameters from fitting a model of the form: \texttt{ tbabs * (ezdiskbb + cutoffpl)}. The parameters from top to bottom are: blackbody normalization (photons cm$^{-2}$ s$^{-1}$) and temperature (keV), powerlaw photon index and normalization (photons cm$^{-2}$ s$^{-1}$) and cutoff energy (keV). \emph{Right Column:} Parameters from fitting a model of the form: \texttt{ tbabs * (ezdiskbb + relxill + cutoffpl)}. Parameters are similar to the left column with the addition of normalization of \texttt{relxill} (photons cm$^{-2}$ s$^{-1}$). }
\label{fig:nu_fit_0_1s}
\end{figure}

The results of the changes in model parameters are shown in the left column of Figure \ref{fig:nu_fit_0_1s}. Five parameters are variable, the temperature ($T_{\rm bb}$) and normalization ($N_{\rm bb}$) of the multicolor blackbody, the photon index ($\Gamma$), high energy cutoff ($E_{\rm c}$) and normalization of the powerlaw ($N_{\rm po}$). The top panel shows the total count rate in the 3--79 keV band. The oscillation is characterized by a slow rise in the blackbody flux at roughly constant temperature, then just prior to the burst peak, the blackbody normalization drops and the temperature rises sharply before dropping again. The other components ($\Gamma$, $N_{\rm po}$, $E_{\rm c}$) track each other, having roughly a constant value before dropping when the blackbody normalization peaks and then rising again to the `normal' value.

There are a few points to note here. The variations in the parameters other than the multicolor blackbody are different from those in \cite{2011ApJ...737...69N} mainly because we use a different model. In that work, the high energy data is described with the \texttt{simpl} model, which takes a fraction of the disk photons and scatters them to high energies \citep{2009PASP..121.1279S}. If we use the same model, namely \texttt{tbabs*highecut*simpl*ezdiskbb}, we obtain results that are similar to \cite{2011ApJ...737...69N}. This highlights the fact that parameters such as $\Gamma, N_{\rm po}, E_c$ depend on the model used in this phenomenological description.

It should also be noted that reflection from the Compton hump is not modeled. Its effect can be seen in the small values of $E_{\rm c}$, which are mostly driven by the spectral curvature due to the Compton reflection hump rather than a real cutoff in the Comptonization component. The value of $N_{\rm po}$ (or $f_{sc}$ in \citealt{2011ApJ...737...69N}) appears higher to compensate for the strong unmodeled reflection hump. To model the full reflection spectrum, we use the model fitted to the average spectrum in section \ref{sec:avg_spec}. To start, all reflection parameters are fixed at their average values. Only the flux (i.e. normalization) is allowed to change (plus $\Gamma$ and $E_{\rm c}$ which are linked to the powerlaw component). The results are plotted in the right column of Figure \ref{fig:nu_fit_0_1s}. We note that the inclusion of reflection always provides a significant improvement in the fit regardless of the continuum model employed.

\subsubsection{Full Modeling with Reflection}

Including the full modeling of reflection has several effects. First, the high energy cutoff increases, as the curvature at $\sim30$ keV is now accounted for by the Compton hump. The cutoff value oscillates between 200 keV when the total flux peaks ($\phi\sim0.9-1.1$) and being unconstrained when the total flux drops. The powerlaw is now steeper compared to the case without reflection because again the hard excess is modeled by the Compton hump. $N_{\rm ref}$ and $N_{\rm po}$ oscillate almost out of phase, with $N_{\rm ref}$ tracking the blackbody disk flux. The powerlaw disappears between $\phi=0.8-1.2$, leaving the hard energy spectrum modeled only by reflection. These latter effects have not been seen before. It is only now, with \nustar data, that the details of the reflection spectrum variability can tracked.

\begin{figure}
\centering
 \includegraphics[width=240pt,clip ]{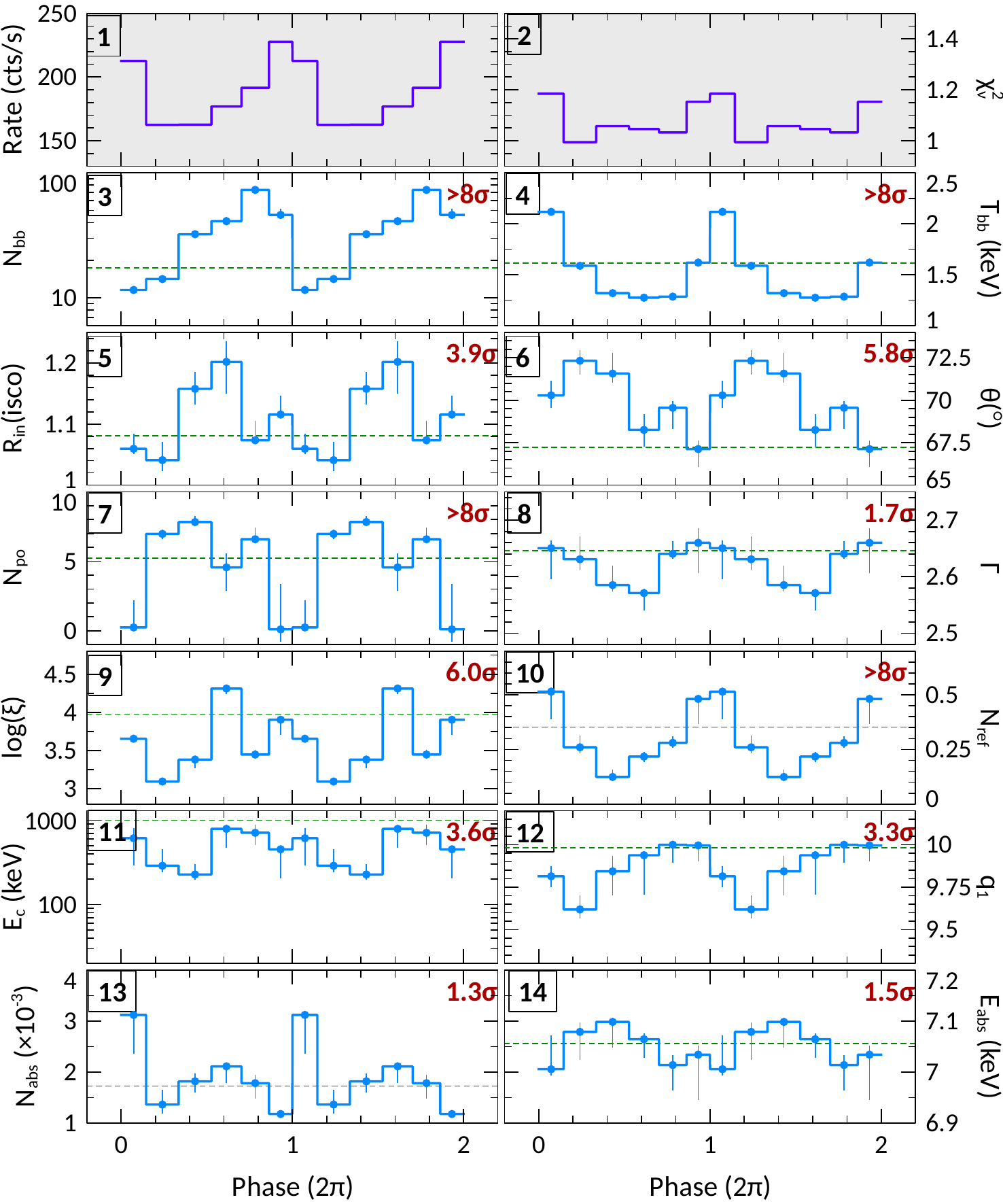}
\caption{Parameter fits from modeling the phase-resolved spectra with the model \texttt{ tbabs * (ezdiskbb + relxill + cutoffpl + gaussian)}. Similar to figure \ref{fig:nu_fit_0_1s}-right but now allowing parameters from the \texttt{relxill} model to vary. Panels 1 and 2 show the total count rate and $\chi^2$ goodness of fit statistic. The remaining parameters are: 3: blackbody normalization, 4: blackbody temperature. 5: Inner disk radius in units of the innermost stable circular orbit of a maximally-spinning black hole. 6: Disk inclination. 7: Powerlaw normalization. 8: Photon index. 9: Disk ionization. 10: normalization of \texttt{relxill}. 11: Cutoff energy of the powerlaw and reflection. 12: Inner emissivity index. 13: Absorbed flux from the narrow absorption line. 14: Energy of the narrow absorption line. The text in the top-right corner of every panel shows the significance of the variation computed through an F-test comparing models where the parameter is allowed and not allowed to change between phases.}
\label{fig:nu_fit_1}
\end{figure}

The bottom panels of Figure \ref{fig:nu_fit_0_1s} show the $\chi^2$ goodness of fit statistic from the model fitting. The addition of the reflection component (right column vs left column plots) makes a significant improvement in the spectral modeling. Note that the reduced $\chi^2$ for most phases is above 1.2, indicating that the other parameters in the reflection spectrum are also changing with phase. Therefore, we test for variability of all reflection parameters except for the iron abundance, emissivity break radius and outer index. The latter two parameters had large uncertainties in the average spectrum and are not constrained in the phase spectra. The results are shown in Figure \ref{fig:nu_fit_1} for six spectral phases. We used six phase bins to allow enough signal in the spectra to search for changes in the relativistic reflection parameters. The phase bins were chosen to have roughly equal numbers of counts per spectrum, and the individual spectra are statistically independent. The count rate (showing the pulse profile) and the goodness of fit statistic are also shown in the top two panels. The significance of parameter changes is calculated using the F-test, comparing models where the parameter of interest is fixed at the average (while all other parameters are allowed to vary) value with models where the parameter changes with phase.

Several points can be noted here:
\begin{enumerate}
\item There is significant improvement in the fits when the reflection parameters are allowed to change with phase (panel 2).
\item The blackbody disk parameters and the normalization of both the powerlaw and \texttt{relxill} are similar to those in Figure \ref{fig:nu_fit_0_1s} (panels 3, 4, 7 and 10).
\item $\Gamma$ is consistent with a constant (panel 8). The changes seen in Figure \ref{fig:nu_fit_0_1s} are now better modeled with ionization changes in the reflection component (panel 9).
\item The ionization parameter $\xi$ and inclination $\theta$ of the disk as inferred from the reflection spectrum change significantly as a function of phase (panels 6 and 9).
\item The inner radius $R_{\rm in}$ also changes significantly (panel 5) and seems to oscillate in phase with the normalization of the blackbody component (panel 4). These two parameters are in principle measuring the same quantity. We will discuss this point further below.
\item The reflection flux (panel 10) is highest when the powerlaw flux (panel 7) is lowest. As we discuss in section \ref{sec:desc_ref}, the corona might be changing in size, and the relation between the reflection and Comptonization is a possible signature of strong light-bending effects.
\item There are hints of possible changes in the absorption line parameters, but we defer further discussion to section \ref{sec:chandra_data}, where \chandra data is better suited at studying the absorption and its variability.
\end{enumerate}

\begin{figure}
\centering
 \includegraphics[width=200pt,clip ]{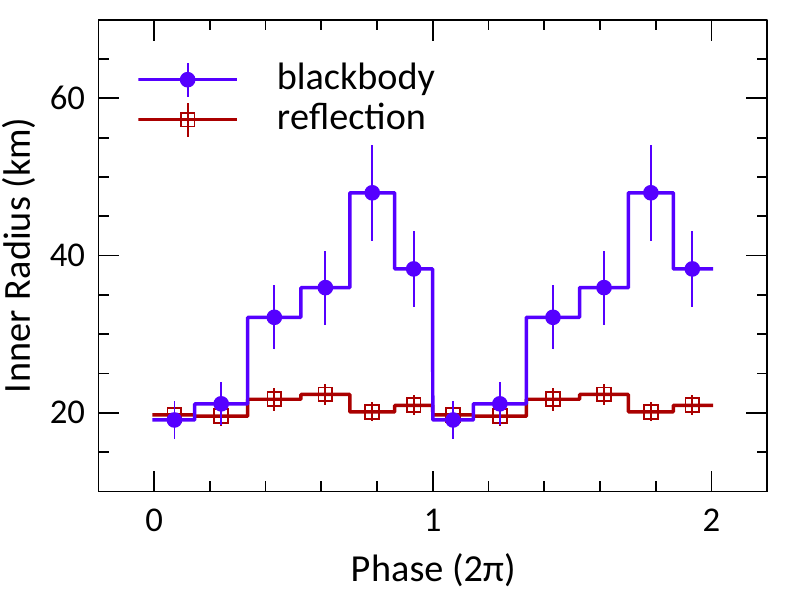}
\caption{Changes in the estimated inner disk radius as inferred from the blackbody emission (blue circles) and from the reflection spectrum (red open squares). The two measurements are clearly not compatible.}
\label{fig:nu_fit1_r}
\end{figure}

$N_{\rm bb}$ is converted to a measure of the inner radius ($R_{\rm bb}$) in units of km using equation \ref{eq:bb}, the inner radius from the reflection spectrum ($R_{\rm ref}$) can also be converted to km assuming again a black hole mass of $M=10.1\pm0.6M_{\odot}$. To calculate $R_{\rm bb}$, we used a color correction factor of $f_c=1.9$ for easy comparison with \cite{2011ApJ...737...69N}. Because the values of $N_{\rm bb}$ do not account for the disk photons that were Comptonized, the naively inferred $R_{\rm bb}$ values are smaller than their true values. We correct for this effect by multiplying the disk normalization with a factor equal to the ratio of total number of photons (disk and powerlaw) to the number of disk photons. A more sophisticated procedure using for example, \texttt{simpl}, is used when discussing detailed model uncertainties in section \ref{sec:desc_ref}. The results are plotted in Figure \ref{fig:nu_fit1_r}. Although the two measures of the inner radius seem to rise and drop at the same phases, the two profiles are very different. $R_{\rm bb}$ changes by a factor of $2.5$ while $R_{\rm ref}$ changes only by a factor of $1.2$. The peak of the phase variations are also different. $R_{\rm bb}$ peaks at $\phi=0.70\pm0.07$ while $R_{\rm ref}$ peaks at $\phi=0.53\pm0.02$ (estimated by fitting a constant plus a Gaussian profile to the phase changes). In terms of the fractional rms variability, $R_{\rm bb}$ has a value of $27\pm9$\% while the corresponding value for $R_{\rm ref}$ is only $5\pm2$\%. It is therefore clear that the two measures of the inner radius are \emph{inconsistent}. The implication of this result and further discussion of uncertainties in the modeling is presented in section \ref{sec:desc_ref}.

\begin{figure}
\centering
 \includegraphics[width=240pt,clip ]{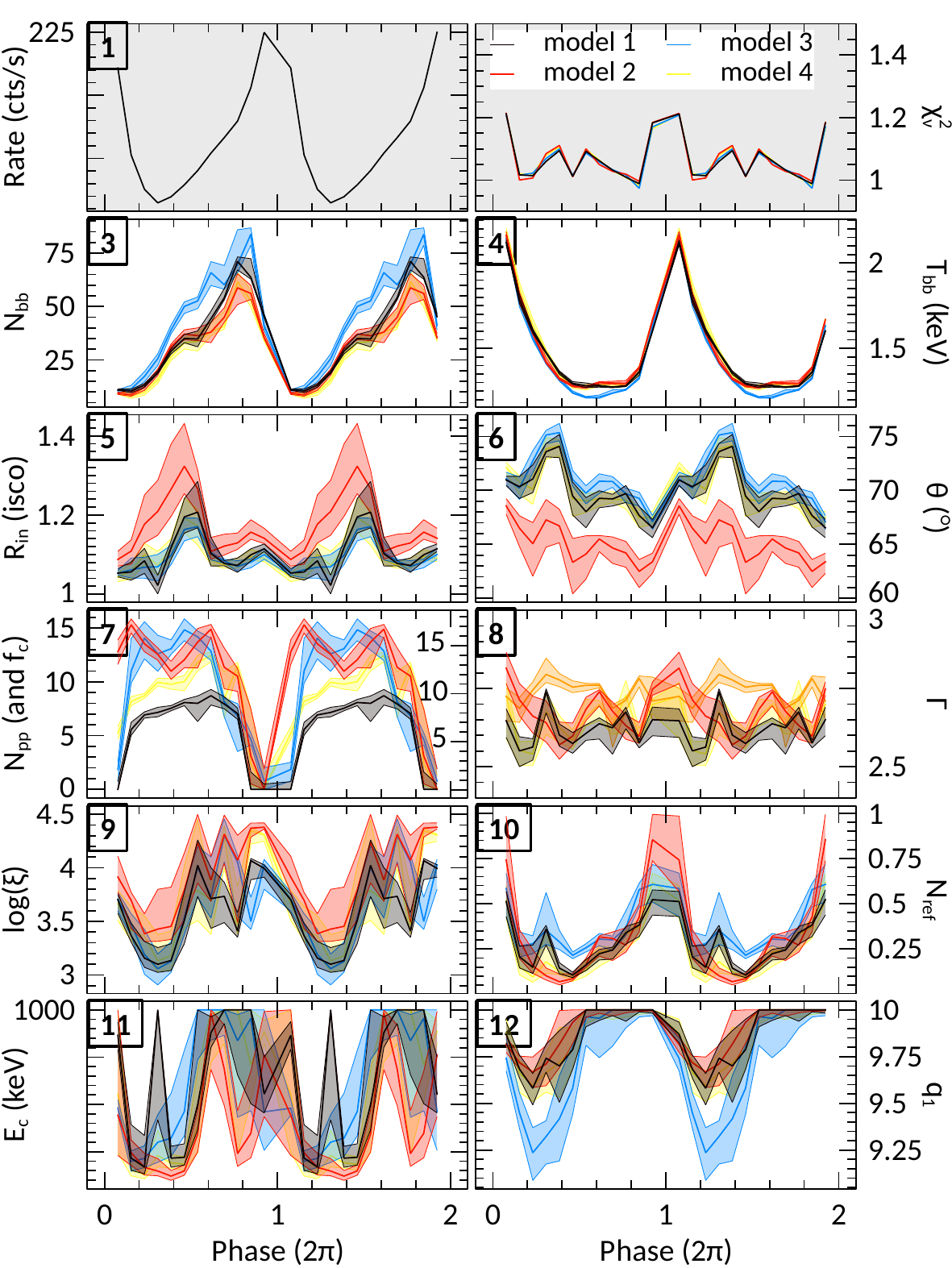}
\caption{Similar to Figure \ref{fig:nu_fit_1} comparing four models discussed in the text. Instead of using six phases bins, the plots here are for 12 phase bins but with an overlap of three bins. This effectively means only four out of the twelve bins are statistically independent. In this way, there is more signal in individual phases and the changes with phases are slightly smoothed out. In panel 7, the right axis if for $N_{\rm po}$ in units of photons cm$^{-2}$ s$^{-1}$ and the left axis is for $f_c$ in $\%$.}
\label{fig:nu_12o_cmb}
\end{figure}

The other important result in Figure \ref{fig:nu_fit_1} is the apparent oscillations in the inner disk inclination, suggesting a possible precession of the disk. The oscillations are at the $10\%$ level, well above the statistical uncertainty in the angle measurements. What is interesting is that the average value is consistent with the inclination of the radio jet of this source \citep{1999MNRAS.304..865F}, while the 10\% changes are consistent with the opening angle of the jet. 

\subsubsection{Alternative Modeling}

Although oscillations in the reflection parameters in Figure \ref{fig:nu_fit_1} are clearly detected, the measurements could be affected by how the continuum is modeled (we will refer to the model in Figure \ref{fig:nu_fit_1} as \texttt{model 1}). To assess the robustness of these measurements, we explore alternative models for hard component as well as additional curvatures in the Comptonization at low energies caused by the high temperature of the seed disk photons.

\begin{figure*}
\centering
 \includegraphics[width=510pt,clip ]{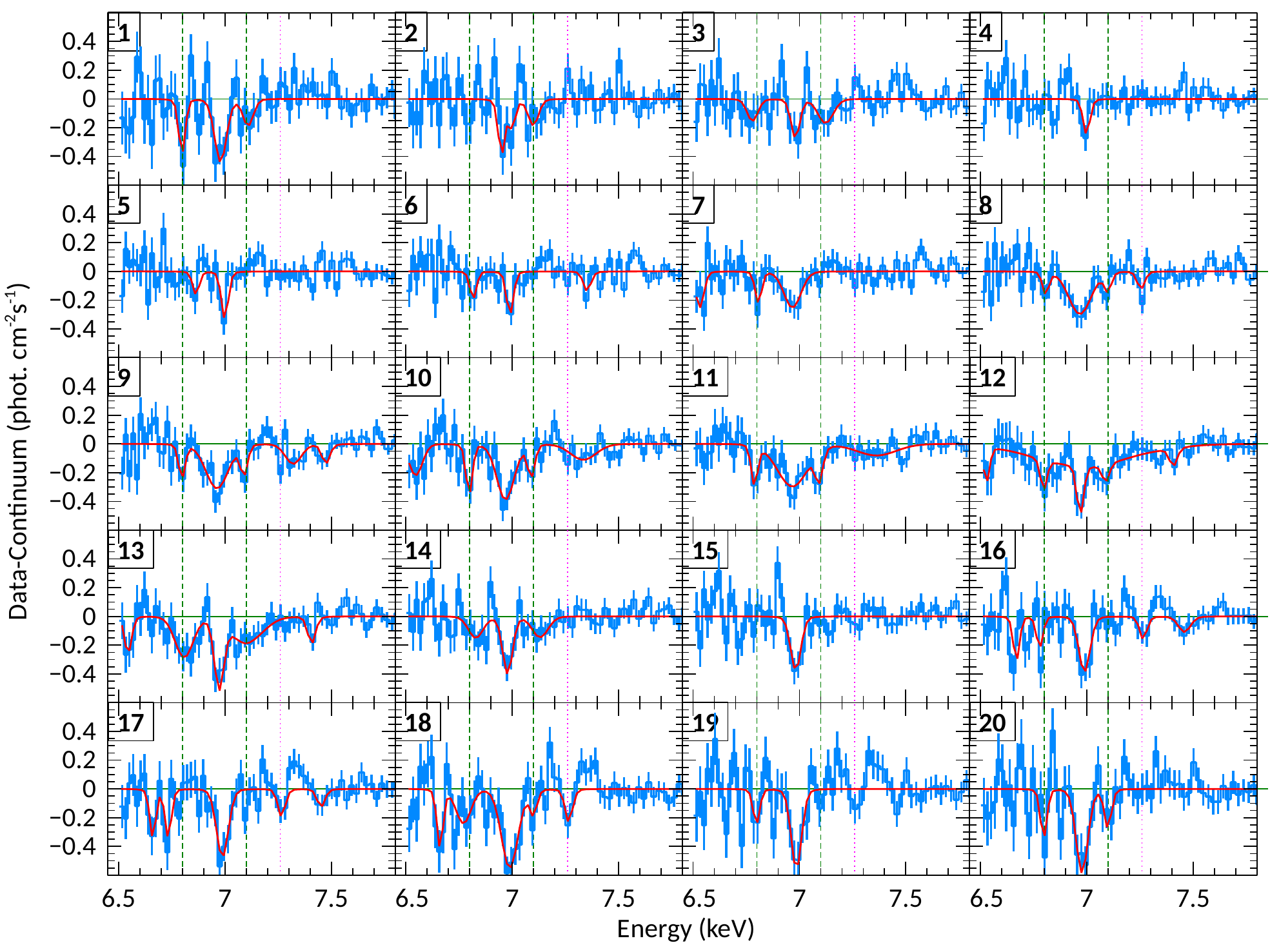}
\caption{Phase-resolved spectra from the \chandra data for 20 phase bins. The overlap in phase is 0.2 (in units of $2\pi$), so that every four bins are fully statistically independent. In other words, panels in different rows are independent and in different columns are not. The spectra are plotted as residuals to a powerlaw model fitted to the 6.5--8 keV band to model the local continuum. The red line represent the continuum with absorption lines that are significant at more than $2.5\sigma$. The vertical dashed green lines mark the position of the pair of lines in the wind component \texttt{c2} with their separation equal to the rest energy separation of a Fe\textsc{xxv}--Fe\textsc{xxvi} pair. The dotted magenta line marks the position of component \texttt{c3}.}
\label{fig:ch_fit5_res}
\end{figure*}

We follow the same procedure as for \texttt{model 1}. First, we fit the average spectrum to obtain the average $N_{\rm H}$, iron abundance, emissivity break radius and the outer emissivity index; then we fit the phase spectra allowing all the remaining model parameters to change. We test three additional models. \texttt{Model 2} includes a low energy cutoff (using \texttt{expabs} in \textsc{xspec}) that affects the powerlaw and reflection models, with the cutoff energy equal to the blackbody temperature. This is an attempt to model the effect of the high temperature of the blackbody which provides the seed photons for the Comptonization process. These Comptonized photons are reflected and the low energy cutoff should also affect the reflection component. Although this effect needs to be included self-consistently in the photoionization modeling of the reflection, the effect can be crudely mimicked by our simple parameterization \citep[see for example Figure 1 in ][]{2013ApJ...768..146G}. So \texttt{model 2} has the form \texttt{tbabs*(gauss+ezdiskbb+expabs* (relxill+cutoffpl))}. In \texttt{model 3}, we model the Comptonization with \texttt{simpl} \citep{2009PASP..121.1279S}. This is an empirical model of Comptonization in which a fraction of the photons in an input seed spectrum is scattered into a powerlaw. The model conserves the number of photons in the Comptonization process and eliminates the divergence of powerlaw models at low energies. We use the multicolor disk spectrum as a seed, similar to \cite{2011ApJ...737...69N}. It turns out however, that because reflection is also important in the spectrum, including the reflection spectrum as part of the seed photons provides a statistically better fit. This is an approximation of the fact that photons reflected from the disk also interact with the electrons producing the Comptonization. \texttt{model 3} therefore has the form: \texttt{tbabs*(gauss+highecut*simpl*(ezdiskbb+relxill))}. In \texttt{model 4}, we fix the iron abundance at solar instead of estimating it from the data. Our measured value in Table \ref{tab:avg_spec} suggests sub-solar abundances, which is different from previous observations \citep{2002ApJ...567.1102L,2013ApJ...775L..45M}. This model therefore tests the effect of the iron abundance on the measured oscillations. The results are shown in Figure \ref{fig:nu_12o_cmb}.

All the models have comparable $\chi^2$ goodness of fit statistics, and many of the features present in Figure \ref{fig:nu_fit_1} are present in the all other models. The effects of the different models is to change the average values of the parameters. One large difference not shown in Figure \ref{fig:nu_12o_cmb} is the change of the best fit average column density when the powerlaw and reflection have a low energy cutoff; the column goes from $N_{\rm h}\sim4.9$ to $\sim3.9 \times10^{22}\, {\rm cm}^{-2}$ as some of the curvature in the data is accounted for by the cutoff and the column density needed is not as high. Nonetheless, it is remarkable that the oscillation seen in Figure \ref{fig:nu_fit_1} is persistent in all models. $\Gamma$ in panel 8 is consistent with constant. Comparing panels 3 and 5, the phase offset present in Figure \ref{fig:nu_fit1_r} is even more apparent. There are no big changes in the fit parameters to account for the $R_{\rm bb}-R_{\rm ref}$ discrepancy. The main conclusion from Figure \ref{fig:nu_12o_cmb} is that \emph{while the exact values for the reflection parameters might depend slightly on the exact model used for the continuum, the oscillations with phase are robust}. The spread in parameters in Figure \ref{fig:nu_12o_cmb} can be taken as a model uncertainty.

\begin{figure*}
\centering
 \includegraphics[width=400pt,clip ]{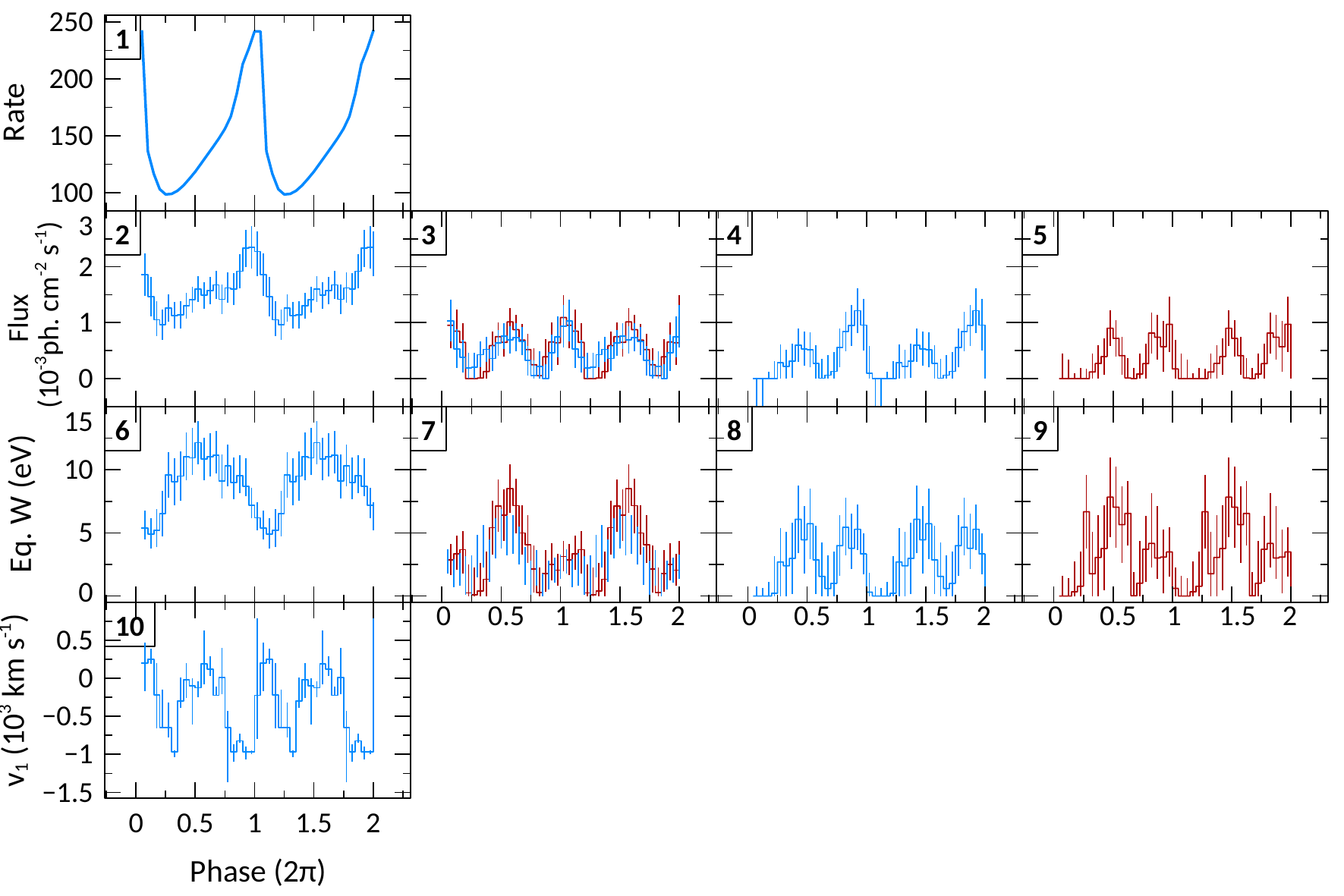}
\caption{Changes of the wind components with oscillation phase. Panel 1 shows the total count rate (counts/s) in the \chandra band. Columns 1, 2, 3 and 4 corresponds to the four wind components \texttt{c1, c2, c3} and \texttt{c4} discussed in the text, respectively. Horizontally, panels 2, 3, 4 and 5 show the absorbed flux (units of $10^{-3}$ photons cm$^{-2}$ s$^{-1}$). The corresponding equivalent widths (in units of eV) are shown in panels 6, 7, 8 and 9, respectively. Panel 10 shows the changes in velocity for \texttt{c1}, which has the highest signal, to allow it to be tracked (in units of $10^{3}$ \kms). For the other components, the velocity was fixed at the average value. 
}
\label{fig:ch_fig9e}
\end{figure*}

\subsection{\chandra} \label{sec:chandra_data}
\chandra also observed \grs on February 23, 2015 (ObsID: 16709). The observation was taken in the Continuous Clocking (CC) Mode, allowing 2.85 msec timing at the expense of one dimension of spatial resolution, which is suitable for timing studies similar to the case at hand. The data were processed using the script \texttt{chandra\_repro}, using \textsc{ciao v. 4.8} and \textsc{caldb v. 4.7.0}. Spectra were reduced in the standard method from \textsc{ciao} documentation. The grating spectra from the positive and negative orders are combined. We only analyze first order spectra as they have the highest signal. The average spectrum in the range of 5--10 keV shows a strong absorption line at $\sim 7$ keV, most likely due to H-like Fe-\textsc{xxvi} at 6.97 keV. There are several emission and absorption lines between 6.4 and 8.5 keV such as the $\sim8.1$ keV line likely due to H-like Ni ($1s-2p$ rest energy at 8.10 keV) and a line at $\sim 6.2$ keV likely due to He-like Mn ($1s^{2}-1s2p$ at 6.18 keV, though this identification is not certain given the absence of Fe-\textsc{xxv} as discussed below) and also weak lines at 6.8, 7.1 and 7.26 keV.  We note here the absence of any strong He-like Fe-\textsc{xxv} line except for a short period during the phase oscillations. As we show below, these lines are variable during the oscillation, and modeling the average spectrum given the variability might not give physical results. Therefore, we proceed directly to phase-resolved spectroscopy.

For timing analysis, light curves with time bins of 0.0685 sec using first and second order grating data were extracted. Barycenter corrections were applied using \texttt{axbary}. GTIs for different phases were extracted similar to the \nustar data in section \ref{sec:nu_phase}. The oscillation profile is found to be very close to that from the \nustar data shown in Figure \ref{fig:nu_waveform}, with small differences due to the different energy bands used to extract the light curves, which do not affect the following results. We extracted 20 phase spectra, where each spectrum is extracted for phase bin widths of $\Delta\phi=0.2$. This ensures that enough signal is available for individual phase spectra, and allows changes to be tracked with phase \citep{2011ApJ...737...69N}. The drawback is that not all the spectra are independent, and in this case, only five are.

Figure \ref{fig:ch_fit5_res} shows residuals of the phase spectra when fitted between 6.5 and 8 keV with a continuum consisting of an absorbed powerlaw and a few Gaussian lines. We start the fits with a powerlaw to model the local continuum and then add Gaussian lines one at a time to model the strongest residuals and stop once the significance of the line drops below $2\sigma$. 

For the purpose of estimating the significance of absorption lines in the following discussion, we use Monte Carlo simulations of fake data based on our best fitting continuum model (absorbed powerlaw). We start with 20 independent faked spectra with a total exposure equal to the total exposure in the observation. From these spectra, another 20 spectra are generated by combining every 4 neighboring spectra. The end result is 20 spectra that are the result of 5 independent spectra, similar to the observation. For every spectrum, we add a Gaussian line to fit the strongest residuals and record the $\Delta\chi^2$ improvement. Repeating this $N=20000$ times gives a distribution of simulated $\Delta\chi^2_s$ with which we compare our observation. A line with an observed $\Delta\chi^2_o$ is detected at $s\%$ significance level when the fraction of simulations with $\Delta\chi^2_s>\Delta\chi^2_o$ is $1-s/100$. The global significance of the line is then obtained using Fisher's combined probability test \citep{10.2307/2681650} using the five independent spectra. We quote the highest number among the 4 possible combinations (though they are comparable), and this significance is also comparable to that obtained directly from the average spectrum. The variability of a parameter is assessed by using the $F$-test, comparing models where the parameter is fixed to the value found in the average spectrum, to models where the parameters are allowed to vary with phase (with the rest of parameters allowed to vary).

\subsubsection{Results}
As can be seen in Figure \ref{fig:ch_fit5_res}, an absorption line at $\sim7$ keV is persistent throughout all phases. We identify the line as being due to Fe-\textsc{xxvi} ($1s-2p$ transition at 6.97 keV). The strength of the line appears to change with phase, being the weakest around phase $\phi\sim0.2$ (panel 4 in Figure \ref{fig:ch_fit5_res}). These changes can be clearly seen in the first column (panels 2, 6 and 10) of Figure \ref{fig:ch_fig9e}, where the absorption due to Fe-\textsc{xxvi} is modeled with a narrow Gaussian line, and the resulting fit parameters are shown as a function of phase, similar to those in section \ref{sec:nu_phase}. The model in this case is a powerlaw plus one absorption line at $\sim7$ keV and the data is fitted between 6.5 and 9 keV. Panel 2 in Figure \ref{fig:ch_fig9e} in particular shows that the absorbed line flux initially decreases, having a minimum at $\phi\sim0.3$, then increases, peaking at $\phi\sim0.9$. 

The absorbed line flux changes are similar to those seen in \cite{2011ApJ...737...69N}, but unlike that observation where a phase delay of 0.92 was observed between the absorbed and total fluxes, we do not see any phase difference. Instead, the two seem to track each other very well, suggesting either a response time that is exactly an integer multiple of the oscillation period, which seems unlikely, or a line response time of $<3$ sec. 
Here, we fixed the line width to the detector resolution. If instead we fit for line width, it is unresolved during most phases except around $\phi=0.2-0.4$ where the width increases to $\sim0.1$ keV. This, however, is not due to an increase in the intrinsic line width, but instead due to the appearance of additional absorption lines around the same energy. This can be clearly seen in panels 8-13 in Figure \ref{fig:ch_fit5_res}, where two lines at 6.8 and 7.1 keV are seen in those phases. The equivalent width of the line is shown in panel 6 in Figure \ref{fig:ch_fig9e} and unlike the absorbed flux, it peaks around $\phi=0.5$ and oscillates between 5 and 12 eV.

If the line is due to Fe-\textsc{xxvi}, then the implied outflow velocity of the absorbing material is $-500\pm100\, {\rm km\, s}^{-1}$ (where the $-$ sign indicates a blueshift), and it oscillates between $250\pm140$ and $-920\pm70\, {\rm km\, s}^{-1}$. The velocity appears to have a period equal to half the main period (i.e. two velocity oscillations within a single flux oscillation).
The existence of the line is highly significant ($>99.99\%$), and it is variable both in flux ($99.8\%$ confidence) and velocity ($96\%$ confidence).
For the purpose of the following discussion, we call this outflow component \texttt{c1}. 
Additional absorption from He-like Fe-\textsc{xxv} at 6.7 keV is seen marginally only in a few phases (panels 16--18 in Figure \ref{fig:ch_fit5_res}). In contrast with the study of \cite{2011ApJ...737...69N}, this suggests the wind properties are very dynamic in nature.

In examining Figure \ref{fig:ch_fit5_res}, we see several additional relatively weak lines that appear to persist in several phases. We note in particular the pair of lines around 6.8 and 7.1 keV whose positions are marked with vertical dashed green lines in Figure \ref{fig:ch_fit5_res}. Their energy separation is consistent with the separation of Fe-\textsc{xxv} and Fe-\textsc{xxvi} at rest energies of 6.7 and 6.97 keV, respectively. If these are due to H and He-like iron, then they require a blueshift of $z=-0.015$ which corresponds to an outflow velocity of $4.5\times10^3\,{\rm km\,s}^{-1}$.
The significance of these lines varies with phase. The global significance of the lines (i.e. the confidence that there is a line in all phases) is $99.96\%$ and $98.5\%$ for the lines at $\sim6.8$ and $\sim7.1$ keV respectively, when taken as independent lines. If they are due to a Fe-\textsc{xxv}--\textsc{xxvi} pair, then significance increases to $>99.99\%$.
The fact that the pair of lines is not seen in all phases suggests variability with phase. To check this, we first find the average energies of these lines by fitting the average spectrum. We find energies of $6.796(3)$ and $7.10(1)$ keV. We then add a Gaussian line to the phase spectra fixing the line energies at their averages and measuring the absorbed line flux and equivalent widths. The results are shown in panels 3 and 7 in Figure \ref{fig:ch_fig9e}. 

As already suggested by Figure \ref{fig:ch_fit5_res}, the absorption from this component is strongest around $\phi=0.1, 0.5$. Although the two lines are fitted independently, their fluxes in panel 3 (and equivalent width in panel 7) vary in phase, strengthening their identification as due to a single outflow component (henceforth \texttt{c2}).
The significance of the peak-to-trough jumps from the first line at $6.796$ keV (shown in blue in panels 3 and 7 Figure \ref{fig:ch_fig9e}) are 94, 92, 95 and 97\% at $\phi\sim0, 0.2, 0.5$ and 0.75, respectively. The significance of the other line is comparable, and by assuming the two lines are produced in the same wind component, the significance is no less that 99.3\%.

There are two additional lines in Figure \ref{fig:ch_fit5_res} that are significant only in some phases, particularly the lines at $\sim7.26$ and $\sim7.4$ keV (components \texttt{c3} and \texttt{c4} respectively). These lines are not very significant globally (87 and 71\% respectively), but they seem to have the same variability as component \texttt{c2} (Figure \ref{fig:ch_fig9e}). If these are due to real lines, both have to vary (with confidence of 85 and 97\%). The energy separation between the \texttt{c3} and \texttt{c4} lines is smaller than the Fe-\textsc{xxv}/Fe\textsc{xxvi} separation, so they are unlikely to be a H- and He-like pair. This could be part of an additional wind component, and if identified with the highly ionized Fe-\textsc{xxvi}, the absorbing material need to be outflowing at velocities of $>1.2\times10^{4}$ and $2\times10^{4}$ km s$^{-1}$ ($0.06\, c$) respectively. These components are comparable in significance to the ultra-fast outflow usually seen in CCD data of AGN \citep{2010A&A...521A..57T}, but here seen in the high resolution \chandra data.

\section{Discussion}
The broad band energy coverage, energy resolution and the high sensitivity of \nustar provide spectral and timing information that allows the diverse phenomena of accreting black holes to be studied in an unprecedented manner. We start our discussion first by directly comparing our results with \citealt{2011ApJ...737...69N} who used \rxte to study the phase spectra of this class of variability. We then discuss the new results inferred from modeling the reflection spectrum.

\subsection{The Continuum}\label{sec:continumm}
The oscillations in our observations are single-peaked. \cite{2012ApJ...750...71N} showed that about a third of all \rxte observations of \grs in the $\rho$-state show a single peak, 40\% show a double peak and the rest show some intermediate shape. The oscillations in fitting parameters are similar to \cite{2011ApJ...737...69N} and \cite{2012A&A...537A..18M} if we use the same models. The shapes, however, depend of the exact model used. One key difference in the modeling is the treatment of reflection. \nustar data not just allows for, but \emph{requires} the inclusion of relativistic reflection. This has a significant effect on other measured parameters. In particular, low energy cutoffs ($<30$ keV) in the Comptonization component are better modeled with a Compton reflection hump rather than a true cutoff. We find the a true cutoff in the continuum is not less than 100 keV at any time during the phase oscillations.

We also find that the photon index of the Comptonization powerlaw remains nearly constant at $\Gamma=2.6$. All observed spectral flattening and softening is caused by the interplay between the different spectral components. This is in line with thermal Comptonization models \citep{1991ApJ...380L..51H}, where the electron temperature adjusts as the seed flux changes to keep the photon index constant.

A constant $\Gamma$ throughout the oscillations implies a constant Compton $y$ parameter, which measures the ratio of heating to cooling \citep{1980A&A....86..121S}. An increase in the soft flux cooling the corona from the disk is matched by an increase in the electron temperature (clear to see in Figure \ref{fig:nu_fit_0_1s}). This is accompanied by an increase in the hard X-ray flux from the corona until $\phi\sim0.9$ where the flux from the corona drops. 
This drop is accompanied by an increase in the reflection fraction, suggesting that the geometry changes, where the disk now sees more of the coronal flux than an observer at infinity. Because the inner radius of the disk is constant, this geometry change could be produced if the corona changes size, becoming smaller and closer to the black hole. Then, due to the strong gravity, more photons hit the disk than reach the outside observer. 

Similar effects are seen in AGN in their dim flux intervals \citep{2008MNRAS.391.2003Z,2012MNRAS.419..116F}. In \grs, the energy density of soft photons that cool the corona increases as the oscillation peaks. The disk also becomes more efficient, and hence more accretion energy is dissipated in the disk rather than in the corona. Additionally, the change in the disk density during the oscillation is likely to change the configuration of the magnetic field, causing changes in the disk-corona energy balance and also in the geometry of the corona. This appears to be similar to that reported recently in \cite{2016arXiv160904592S}, where stronger reflection was reported in the disk-dominated soft state observations of 29 stellar-mass black hole candidates.

\begin{figure}
\centering
 \includegraphics[width=220pt,clip ]{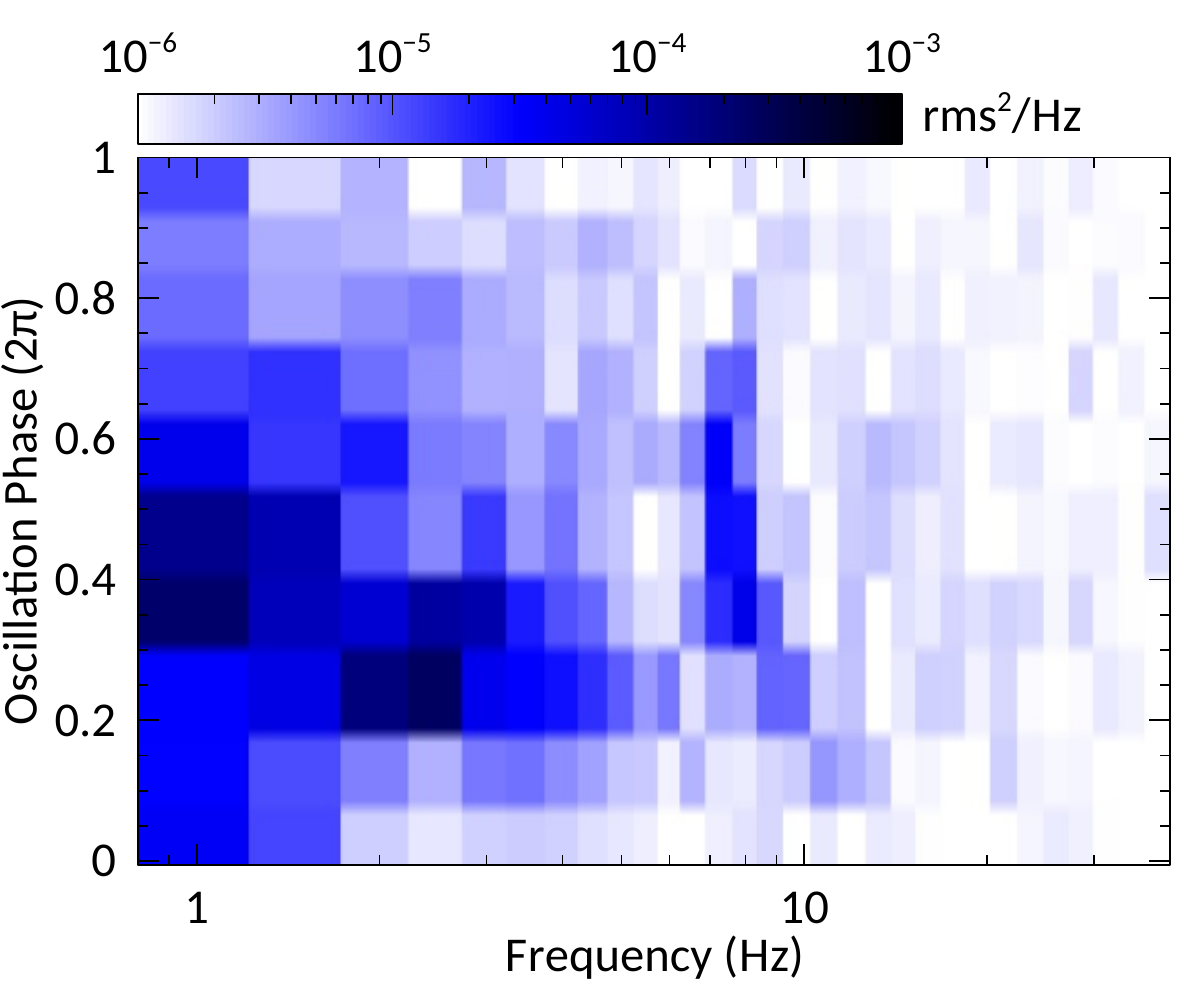}
\caption{Phase-resolved co-spectrum from the \nustar data. The co-spectrum is calculated following \cite{2015ApJ...800..109B} and it is similar to the standard power spectrum but additionally corrects for the non-constant dead-time in \nustar. A low frequency QPO is clearly present between phases 0.2--0.8. These are the phases that our modeling indicate the Comptonization component is present. This component is absent during the rest of the phases where the QPO also disappears.}
\label{fig:nu_cospec}
\end{figure}

 A corona changing in size is further supported by the matching changes in the low frequency QPO at $\sim10$ Hz in the data. QPOs are known to be associated with the Comptonization component \citep{1997ApJ...488L.109B,2000ApJ...531..537S}, being stronger when the contribution of the Comptonization component is high. We show the phase-resolved co-spectra in Figure \ref{fig:nu_cospec}. The co-spectrum is calculated by correlating light curves from the two \nustar modules at different phases. The co-spectrum in this case gives similar results to the standard power spectrum with the advantage of correcting for the non-constant dead-time effects from the detectors \citep{2015ApJ...800..109B}.

The QPO is significantly detected above the noise just below 10 Hz in the phase range $0.2<\phi<0.8$, the same phases where our modeling indicates strong Comptonization component, and it is weaker when the powerlaw component is weakest, \emph{despite the observed hard flux being roughly constant}. The size change for the corona is further supported by the change in the frequency of the QPO. Between phases 0--0.3, the frequency is higher, suggesting a smaller scale size, as we already inferred from the continuum flux changes. This result is in line with the results reported in \cite{2013ApJ...767...44Y}.

\subsection{The Inner Radius of the Disk}\label{sec:desc_ref}
We found that the inner radius of the disk as inferred from the blackbody emission appears to oscillate, similar to early measurements \citep{1997ApJ...488L.109B}. The radius measured from the reflection spectrum on the other hand remains nearly constant. So these two estimates are clearly \emph{not} measuring the same quantity.

This result is robust against the details of the modeling in a sense that the exact values for $R_{\rm bb}$ and $R_{\rm ref}$ might depend on the model slightly, the fact that $R_{\rm bb}$ varies varies by $\sim30\%$ while $R_{\rm ref}$ varies by no more than $\sim5\%$ holds all the time. We tested this by using many alternative modelings including models with a low energy cutoff (to model to Comptonization from high temperature blackbody), models with \texttt{simpl}, models with different assumptions about the reflection parameters and more sophisticated models for the Comptonization (using \texttt{eqpair}). \emph{In all cases, oscillation in $R_{\rm bb}$ is observed while $R_{\rm ref}$ is almost constant.}

It is unlikely that changes in the spectrum conspire to make $R_{\rm ref}$ appear constant when it is not (it needs fine tuning), while it is slightly more likely that a constant $R_{\rm bb}$ appears to change. This arguments along with more grounded theoretical predictions \citep{2000ApJ...535..798N} would suggest that the real inner radius of the disk does not change. The implication is that some of the assumptions (density profile for instance) in the \texttt{ezdiskbb} model (and similar models) might be invalid. 

If we now assume $R_{\rm ref}$ measures the true inner radius of the disk, then the discrepancy between $R_{\rm bb}$ and $R_{\rm ref}$ can be due to an error in the black hole mass (to convert $R_{\rm ref}$ from gravitational to physical units), the distance $D$ or the color correction factor $f_c$ (see equation \ref{eq:bb}). The first two only change the absolute value. The difference in the strength of variability of each parameter implies that the color correction factor ($f_c$) {\it changes} with phase. We have so far assumed the color correction has a constant value of $f_c=1.9$. The fact that $R_{\rm ref}<R_{\rm bb}$ implies $f_c<1.9$. 

\begin{figure}
\centering
 \includegraphics[width=220pt,clip ]{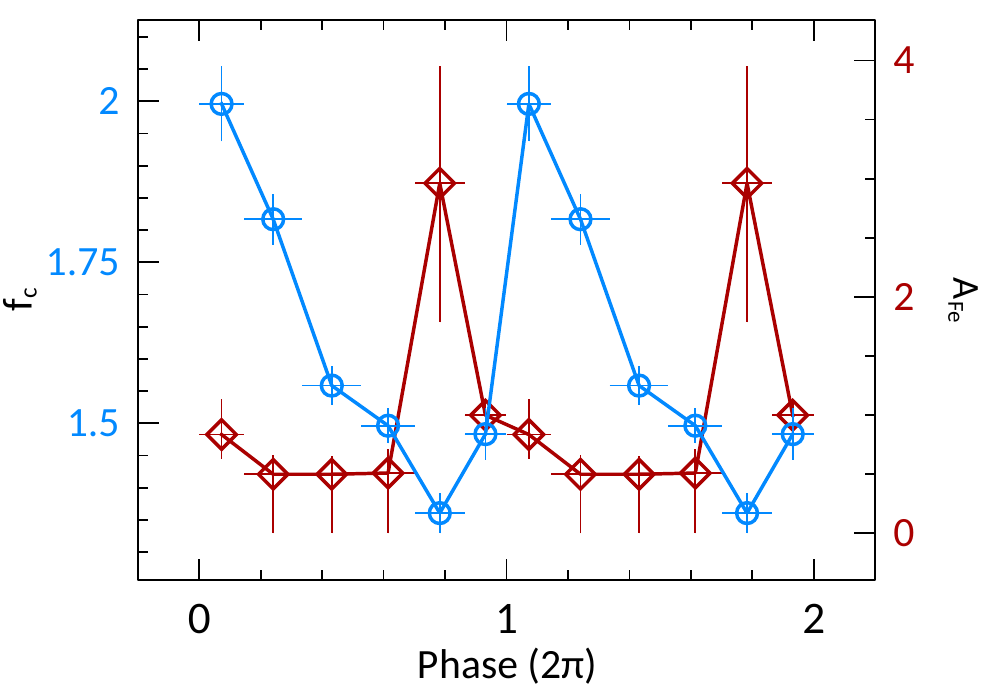}
\caption{Oscillations in the color correction factor $f_c$ (blue circles, left axis) required to make radii measured from the disk blackbody ($R_{\rm bb}$) and reflection ($R_{\rm ref}$) equal. The red diamonds show the oscillation in the iron abundance measured from the reflection spectrum (right axis). The two are anti-correlated.}
\label{fig:nu_f_fe}
\end{figure}

If we calculate the required $f_c$ values that make $R_{\rm bb}=R_{\rm ref}$, we find that it changes between 2 at $\phi=0$ to 1.2 at $\phi=0.9$, as shown in Figure \ref{fig:nu_f_fe}, and a significant softening (i.e. closer to a true blackbody) is required. A change in the disk density and/or composition can be responsible for these changes. A change in the disk surface density is expected from theoretical modeling \citep{2000ApJ...535..798N}, but it is not clear how this can be tested observationally, independently of the blackbody flux. Disk atmosphere models also predict a change in $f_c$ with luminosity \citep{2006ApJ...636L.113S}. However, those are not large enough to explain the changes in $f_c$ observed here. The change in composition can be directly tested given the reflection spectrum. The strength of the iron line in particular is sensitive to the iron abundance. In our modeling, we assumed the iron abundance is fixed at the average. If we allow it to change, we find the variations shown in Figure \ref{fig:nu_f_fe}. First, the iron abundance appears to change between 0.5 to 3 times the solar value (significant at the 95\% confidence level). Second, and most remarkably, the abundance appears to be higher when $f_c$ is lowest. One possible explanation is that the iron content in the disk is changing during the oscillation, possibly by radiation levitation \citep{1996Ap&SS.237..107S}. 

Iron levitation by radiation forces in accretion disk is discussed by \cite{2012ApJ...755...88R}. The ratio of drift to dynamical time-scales for many iron ions drops with temperature, and it is of order 1 or less at the inner regions of the disk where the temperature is $\sim2\times10^7$ K, similar to the case at hand \citep{2012ApJ...755...88R}. Therefore, as the disk temperature and flux increase, radiation pushes iron ions up in the disk causing an increase in the observed iron abundance, and most crucially, {\emph increasing} the opacity in the upper disk atmosphere. This makes the disk radiation appear closer to a true blackbody, and hence the color correction is small. As the disk radiation drops, the force on iron drops and the iron sinks again to the inner layers of the disk. We also note a possible correlation between the abundance and ionization of the disk driven likely by the density and composition changes during the instability.  Although these results are only suggestive and the physics might be different, it is conceivable that similar composition effects could affect the measurements of the inner radius of the disk in the canonical states of black hole binaries.

\subsection{Accretion Physics and the Nature of Oscillations}
In the modeling of \cite{2000ApJ...535..798N} and \cite{2000ApJ...542L..33J}, although the inner radius was \emph{assumed} constant, blackbody emission from the inner regions during the outburst changes in a way that mimics a changing inner disk radius \citep[see Figure 11 in][]{2000ApJ...535..798N}. The radius changes very little during the oscillations while changes in the disk as the instability waves propagate suppress the blackbody emission from the inner regions. The wave propagation changes the density, and may contribute to the metal abundance changes discussed earlier. Our results seem to confirm observationally the validity of such assumptions.

One additional observation we note here is the apparent change in the inclination of the inner disk. This might indicate a precession, however it is not clear if the thermal instability can cause the disk to precess. One change in the geometry of the disk that models predict is a change in the $H/R$ in some parts of the disk as the density/heating waves propagate \citep{2000ApJ...542L..33J,2001MNRAS.328...36S}. The steep emissivity profiles inferred from the spectra suggest that the reflection spectrum is produced in the inner few gravitational radii. A smooth change in the scale height with radius will appear as a bulge in the disk extending over several to tens of gravitational radii, and might cause the apparent inclination of the disk to change. As the bulge moves out before disappearing as the instability saturates \citep[see Figure 1 in ][for instance]{2001MNRAS.328...36S}, the apparent inclination will oscillate. Using the simulations of \cite{2001MNRAS.328...36S} and \cite{2000ApJ...542L..33J} as a guide, $H/R$ can reach $H/R\sim0.1-0.2$ at radii of $\sim10-20 r_g$ (these are local temporary changes). This can cause an inclination change of $\Delta\theta\sim6-12^{\circ}$, which is similar to what we observe (panel 6 in Figure \ref{fig:nu_fit_1}). 

These $H/R$ changes can be also invoked to explain the small changes in $R_{\rm ref}$ inferred from reflection. Although these changes are very small compared to changes in $R_{\rm bb}$, they are nonetheless significant in the data. This can be caused by the bulge eclipsing parts of the very inner disk during parts of the oscillations, giving the appearance of small changes in radius. This a natural consequence of the highly inclined disk we observe.

\subsection{Disk Winds}
\cite{2011ApJ...737...69N} analyzed \chandra data similar to ours probing the same state. It it worth making a direct comparison, as this will allow the wind properties that are associated with the $\rho$-state to be separated from those transient events that are specific to individual observations. First, we do not see any strong absorption from Fe\textsc{xxv} at 6.7 keV at rest or at low blueshifts during the oscillation, in contrast to \cite{2011ApJ...737...69N}. Absorption from Fe-\textsc{xxvi} at 6.97 keV is present, and the absorbed flux in the line tracks the total flux. Given that the source luminosity and the luminosity change with phase are similar between the two observations, the absence of the Fe-\textsc{xxv} line implies the wind has a different ionization and hence either the distance or the density have changed between this observation and that analyzed by \cite{2011ApJ...737...69N}. The higher ionization in the current observation suggests that the outflowing gas has either a \emph{lower} density or is located \emph{closer} to the ionizing source or both. Given our radius estimates discussed next suggests that the latter (i.e. smaller distance) is likely.

The phase match between the total illuminating flux and the changes in the absorbed flux, and the absence of any delay down to the phase resolution allowed by the data ($t_{\rm period}/n_{\rm phases}\sim2$ sec) implies that light travel time effects longer than 2 sec are ruled out, putting the wind at $R_{\rm wind}<6\times10^{10}{\rm cm}$. At these distances, given that we observed a wind velocity of $v_{\rm wind}=500$ \kms, the estimated dynamical timescale of the wind is $t_{\rm dyn}\sim R_{\rm wind}/v_{\rm wind}\sim1000 $ sec. However, we observe changes in the wind on time scales of $\sim 4$ sec (e.g. changes in velocity in panel 10 in Figure \ref{fig:ch_fig9e}). This suggests that the wind is launched from a region that is much smaller ($R_{\rm wind}\sim3\times10^8{\rm cm}$).

The observed velocity changes do not seem to be directly associated with the absorbed flux, as there appears to be \emph{two} cycles of change in velocity within a single cycle of flux oscillation. The same two cycles are also seen in the flux of at least other two wind components. If light - through flux changes - is the only communication channel between the source and the wind, and the wind sees the same continuum we see, then it is generally difficult producing the half-period oscillation in the velocity when all model parameters from the spectral modeling have a period equal to the total oscillation period.
Even if a distant wind sees a different continuum, which could happens if, for instance, a central compact source is eclipsed by a flaring or bulging disk, then it is still not clear how a half-period oscillation in the wind velocity can be produced. If, however, the wind is launched from small radii, where geometrical changes associated with the limit cycle are taking place, then such half-period oscillations can be envisaged.

\begin{figure}
\centering
 \includegraphics[width=220pt,clip ]{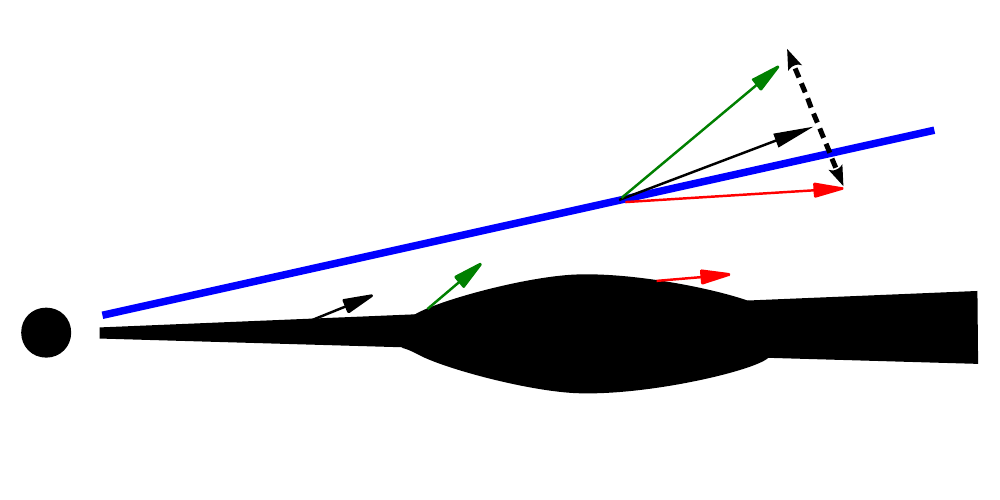}
\caption{An illustration of the possible geometry change and its effect on the wind. The change in $H/R$ during the instability causes the line-of-sight wind velocity to change producing the half-period oscillation in the velocity of the first component. The blue line is our viewing sight-line. Thee three arrows on the disk show the direction of the wind that changes due to the changes in $H/R$. Their angle to the local disk normal is constant. A single bulge moving through the wind-emitting region produces a half-period oscillation in the line-of-sight wind velocity. Two minima are produced at the extremes of $H/R$ changes, and one maximum is produced when the wind is in our line-of-sight.}
\label{fig:wind_illustration}
\end{figure}

For instance, the change in $H/R$ predicted by models and supported by the oscillations in the inclination of the inner disk can produce the half-period oscillations of the velocity. In the simplest picture, a bulge moving in the disk during the instability will cause the line-of-sight velocities of an otherwise constant wind to appear to oscillate (see illustration in figure \ref{fig:wind_illustration}). The half-period oscillation is produced if the wind is launched with an angle close to our line-of-sight. Two minima in the line-of-sight velocity are produced by the maximum $H/R$ changes, and a maximum is produced when the wind direction crosses our line-of-sight. Although it is not clear how far out in the disk the H/R changes might extend to, simulations \citep{2001MNRAS.328...36S} suggest that they could reach a few 100 $r_g$, consistent with wind launch region. Our highly inclined view of \grs ($\theta\sim67^{\circ}$) suggests that the wind has be launched at a comparable angle. An equatorial wind is consistent with many studies of winds in black hole binaries \citep{2012ApJ...746L..20K,2012MNRAS.422L..11P}. This picture explains the half-period oscillation only, not the velocity values, as the velocity in this case is not expected to reach zero during the oscillation. The fact that we observed the velocity reaching zero suggests that there must be intrinsic velocity changes or additional mechanisms that slows the wind when its direction changes, possible by colliding with outflowing material not affected by the geometry change.

It is therefore clear given the oscillation profiles of the velocity of the first wind component (\texttt{c1}), and the absorbed flux and column of the other components, that some \emph{changes in the geometry} in the wind are required to explain the half-period oscillations. The shortest time-scale that can be linked to the limit cycle oscillation is the thermal time-scale: $t_{\rm th} = 5\times10^{-5}\alpha^{-1} r^{1.5}$ sec, where $\alpha$ is the viscosity parameter of the standard disk, $r$ is the disk radius in units $r_g$, and we have taken $M=10 M_{\odot}$ \citep{2002apa..book.....F}. If we equate this to the half-period oscillation suggested by the change in wind parameters (25 sec), and assume $\alpha=[0.01-0.1]$, we obtain a radius of $R_{\rm wind}=[290-1300] r_g$($=[4-18]\times10^{8}$ cm). This is consistent with our earlier dynamical estimate. Given the observed high ionization $log\xi>4.7$ inferred from the absence of the Fe\textsc{xxv} line, and given the luminosity of the source that reaches $10^{38}\, {\rm erg} {\rm s}^{-1}$, the wind density is estimated to be $n_{\rm wind}=[0.6-10]\times10^{15}\, {\rm cm}^{-3}$. These measurements are consistent with those obtained from the profiles of absorption and emission lines from an observation in an extended soft state of \grs \citep{2016arXiv160301474M}. These estimates should be taken as upper limits on the density given the lower limit on the ionization we observe.

In order to estimate the physical properties of the wind self-consistently, we model the absorption lines from the H-, and He-like Fe using the photoionzation code \textsc{xstar} \citep{2000HEAD....5.2703K}. We calculate the physical conditions in a gas illuminated by the spectra observed with \nustar. We assume the gas density to be $n=10^{15} {\rm cm}^{-3}$ as inferred from the estimated wind launch radius. We assume a covering factor of 0.5 and fit directly for the equivalent Hydrogen column density ($N_{\rm H}$), ionization outflow velocity of the wind. We find that for component \texttt{c1}, $N_{\rm H}$ has an oscillation profile similar to the equivalent width shown in panel 6 of Figure \ref{fig:ch_fig9e}, with the value oscillating between $5\times10^{22}$ and $5\times10^{23} {\rm cm}^{-2}$. The ionization parameter of \texttt{c1} has mostly a lower limit of $log\xi=4.7$ as inferred from the absence of Fe-\textsc{xxv}, and drops to $log\xi=4$ around $\phi=0.8$ where a weak Fe-\textsc{xxv} appears in the spectrum (panel 15 in Figure \ref{fig:ch_fig9e}). It should be noted here that it is difficult to distinguish changes in $N_{\rm H}$ from changes in $\xi$ at these high ionizations, as the only source of information is the Fe-\textsc{xxvi} line. For the other components, the oscillation profiles in $N_{\rm H}$ are again similar to those of the equivalent width shown in Figure \ref{fig:ch_fig9e}, which the values oscillating between $5\times10^{21}$ and $5\times10^{22} {\rm cm}^{-2}$. The ionization parameters oscillate with a lower limit of $log\xi\sim4$.

The mass outflow rate in the wind can be estimated using a spherical approximation. In this case $\dot{M}_{\rm wind}=C_{\rm V} \Omega m n R_{\rm wind}^2 v$, where $v$ is the velocity of the wind ($v\sim500$\kms for \texttt{c1}), $m$ is the average ion mass per Hydrogen atom ($m=2.4\times10^{-24}$ g), $\Omega$ is the total opening angle of the wind (assumed to be $\sim\pi$) and $C_{\rm V}$ is the volume filling factor. We find that for \texttt{c1}, $\dot{M}_{\rm wind}\sim0.04-10\times10^{18}$ $C_{\rm V}$ g s$^{-1}$. For \texttt{c4}, which has the highest velocity, $\dot{M}_{\rm wind}\sim2\times10^{19}$ $C_{\rm V}$ g s$^{-1}$ for the range of measured densities. 
$C_{\rm V}\sim dR/R_{\rm wind}$ with $dR\sim N_{\rm H}/n=[0.5-5]\times10^8$ cm, giving $C_{\rm V}\sim0.1-0.3$.
Given the less than unity duty cycle inferred from the variability, the mass outflow rate is roughly $\sim0.1-15$ times the mass accretion rate, and represents about 0.01--0.3\% of the radiative luminosity of the source.

We note that the only other object that shows similar variability to \grs is IGR J17091-3624 \citep{2011ApJ...742L..17A}. Most of the patterns reported here are expected to be present there too. Future observations might be able to test this.

\section{Summary}
We have presented phase-resolved spectroscopy of the heartbeats state of \grs using data from \nustar and \chandra. This is the first time the details of the reflection spectrum are studied during the remarkable oscillations of \grs. The main results from the analysis are as follows:
\begin{itemize}
\item The extent of the inner disk inferred from the reflection spectrum changes very little during the oscillations regardless of the model used for the continuum. This is in contrast with the radius measurements inferred from the blackbody emission. The two measurements may be reconciled if the disk density changes as suggested by theoretical modeling and/or the temperature color correction factor of the blackbody changes significantly during the oscillations. Changes in the composition of the disk atmosphere may be responsible for the oscillations in the color correction factor. We observed suggestive evidence that the iron abundance changes during the oscillations, supporting this possibility.
\item The inferred black hole spin and disk inclination are similar to those measured the plateau state of \grs.
\item The measured disk inclination matches the inclination of the axis of the radio jet remarkably well despite the two very different measurement methods. The inclination inferred from the reflection spectrum oscillates by about $10^{\circ}$ in the heartbeats state, which is comparable to the opening angle of the jet. The latter connection could however be just a coincidence given the fact that jets are mostly observed in the low C state in \grs where no oscillations are observed.
\item Changes in the low frequency QPO during the oscillations and reflection fraction indicate that the corona collapses in size at the peak of the oscillation.
\item The simultaneous \chandra data show the presence of at two wind components that change with oscillation phase with velocities ranging from 500 \kms\, to $5\times10^{3}$ \kms\,, and weak evidence for higher velocity components with velocity of $0.06\, c$ which are comparable to the ultra fast outflows seen in AGN with CCD resolution but seen here at high resolution (see figures \ref{fig:ch_fit5_res} and \ref{fig:ch_fig9e}). 
\item An upper limit of the wind response time of 2 sec inferred from the absorbed flux oscillations (panel 2 in Figure \ref{fig:ch_fig9e}) puts an upper limit of the wind launch radius of $6\times10^{10}\, {\rm cm}$.
\item Oscillations of half the main oscillation period in the wind velocity of the first component (panel 10 in Figure \ref{fig:ch_fig9e}) and the strength of the wind in other components (panels 7, 8 and 9 in Figure \ref{fig:ch_fig9e}) further constrain the wind to be at $\sim290-1300\, r_g$.
\item The wind carries about 0.1--1\% of the radiative luminosity of the source, and a mass outflow rate at least comparable to the accretion rate.
\end{itemize}

\section*{Acknowledgment}
This work made use of data from the {\em NuSTAR} mission, a project led by the California Institute of Technology, managed by the Jet Propulsion Laboratory, and funded by the National Aeronautics and Space Administration. 
This work is also based on observations made by the Chandra X-ray Observatory.

\bibliographystyle{astron}
\bibliography{apj}

\end{document}